\documentclass[a4paper,twocolumn,11pt,accepted=2023-10-03]{quantumarticle}
\pdfoutput=1
\usepackage[utf8]{inputenc}
\usepackage[english]{babel}
\usepackage[T1]{fontenc}
\usepackage{amsmath}
\usepackage{amssymb}
\usepackage{hyperref}
\usepackage{braket}
\usepackage[numbers,sort&compress]{natbib}

\begin{document}
\title{Fermion-qudit quantum processors for simulating lattice gauge theories with matter}

\author{Torsten V. Zache}
\email{torsten.zache@uibk.ac.at}
\affil{Institute for Theoretical Physics, University of Innsbruck, 6020 Innsbruck, Austria}
\affil{Institute for Quantum Optics and Quantum Information of the Austrian Academy of Sciences,
6020 Innsbruck, Austria}
\affil{These authors contributed equally to this work.}

\author{Daniel Gonz\'{a}lez-Cuadra}
\email{daniel.gonzalez-cuadra@uibk.ac.at}
\affil{Institute for Theoretical Physics, University of Innsbruck, 6020 Innsbruck, Austria}
\affil{Institute for Quantum Optics and Quantum Information of the Austrian Academy of Sciences,
6020 Innsbruck, Austria}
\affil{These authors contributed equally to this work.}

\author{Peter Zoller}
\affil{Institute for Theoretical Physics, University of Innsbruck, 6020 Innsbruck, Austria}
\affil{Institute for Quantum Optics and Quantum Information of the Austrian Academy of Sciences, 6020 Innsbruck, Austria}

\begin{abstract}
	Simulating the real-time dynamics of lattice gauge theories, underlying the Standard Model of particle physics, is a notoriously difficult problem where quantum simulators can provide a practical advantage over classical approaches. In this work, we present a complete Rydberg-based architecture, co-designed to digitally simulate the dynamics of general gauge theories coupled to matter fields in a hardware-efficient manner. Ref.~\cite{Gonzalez_2022} showed how a qudit processor, where non-abelian gauge fields are locally encoded and time-evolved, considerably reduces the required simulation resources compared to standard qubit-based quantum computers. Here we integrate the latter with a recently introduced fermionic quantum processor~\cite{Gonzalez_2023}, where fermionic statistics are accounted for at the hardware level, allowing us to construct quantum circuits that preserve the locality of the gauge-matter interactions. We exemplify the flexibility of such a fermion-qudit processor by focusing on two paradigmatic high-energy phenomena. First, we present a resource-efficient protocol to simulate the Abelian-Higgs model, where the dynamics of confinement and string breaking can be investigated. Then, we show how to prepare hadrons made up of fermionic matter constituents bound by non-abelian gauge fields, and show how to extract the corresponding hadronic tensor. In both cases, we estimate the required resources, showing how quantum devices can be used to calculate experimentally-relevant quantities in particle physics.
\end{abstract}

\maketitle

\section{Introduction}
According to the Standard Model of particle physics, interactions between elementary particles, such as electrons and quarks, are mediated by gauge bosons like photons or gluons~\cite{weinberg1995quantum}. These interactions are described by quantum field theories that are invariant under local (or gauge) transformations, commonly known as gauge theories. Due to the non-perturbative nature of many relevant high-energy phenomena, especially those described by non-abelian gauge theories like quantum chromodynamics (QCD), numerical methods are necessary to extract predictions from first principles, which can then be compared to experimental results. A prime example of this is quark confinement, which is responsible for the stability of hadronic matter at low temperatures and densities. Many properties of hadrons, which are bound states of quarks and gluons, such as protons or neutrons, have been predicted and later experimentally confirmed using lattice gauge theories (LGT), in which space and time are first discretized to enable the use of Monte Carlo methods~\cite{montvay1997quantum, Aoki_2019}.

Despite its success, Monte Carlo methods sometimes encounter a severe sign problem~\cite{Troyer_2005}. This problem prevents the study of fermionic fields at finite chemical potentials, as well as the real-time dynamics of lattice gauge theories (LGTs). Therefore, finding alternatives to standard methods is crucial, as many open problems in particle physics require calculations in regimes where the latter fail. For example, the phase diagram of quantum chromodynamics (QCD) is only known at small baryon densities, and we still lack a quantitative understanding of phases such as the quark-gluon plasma (QGP), where quarks become deconfined, as well as the nature of the corresponding confined-deconfined transition~\cite{Brambilla_2014}. Similarly, many questions remain about the non-equilibrium properties of QCD~\cite{berges2021qcd}. Of particular relevance is the thermalization of the QGP after being produced in heavy-ion colliders, where extremely high temperatures, similar to those in the early universe, are reached, as well as its hadronization at lower temperatures.

Recently, quantum simulators have emerged as a promising alternative to classical methods~\cite{Wiese_2013, Zohar_2015, Dalmonte_2016, Banuls_2020, Aidelsburger_2022, Zohar_2022, dimeglio_2023}, overcoming several limitations as demonstrated in various experiments studying equilibrium and non-equilibrium properties of LGTs using analog and digital approaches~\cite{Martinez_2016, Schweizer_2019, Kokail_2019, Mil_2020, Yang_2020, Zhou_2021, Nguyen_2021}. Analog quantum simulators~\cite{Cirac_2012, Georgescu_2014} involve engineering quantum systems such as ultracold atoms~\cite{Gross_2017, Browaeys_2020} or trapped ions~\cite{Blatt_2012, Monroe_2021} to behave under the same Hamiltonian as the simulated system, allowing for large system sizes, but they are limited in the complexity of simulated interactions. Digital quantum simulators have universal control and can target non-abelian gauge theories in higher dimensions~\cite{Byrnes_2006, Lamm_2019, Alexandru_2019, Ji_2020, Mathis_2020, Kaplan_2020, Brower_2020, Shaw_2020, Kclo_2020, Kclo_2021, Alexandru_2021, Haase_2021, Bauer_2021, Kan_2021, Davoudi_2021b,klco2022standard}, and various implementations have been proposed for experimental platforms~\cite{Muschik_2017, Paulson_2021, Davoudi_2021a, Tagliacozzo_2013a,Tagliacozzo_2013b,Zohar_2017a,Zohar_2017b,Bender_2018, Mezzacapo_2015, Klco_2018, Atas_2021, Armon_2021}. However, current quantum computers are limited in system size and circuit depth~\cite{Preskill_2018}, and to perform quantum simulations with practical advantage~\cite{Daley_2022}, quantum hardware and software need to be co-designed and tailored to the specific problem at hand to manage resources.

In Ref.~\cite{Gonzalez_2022}, we introduced a Rydberg-based architecture where {\em qudits} are encoded as internal states of atoms trapped in optical tweezers~\cite{Ebadi_2021, Scholl_2021, Kaufman_2021}, and a universal set of qudit gates was designed using laser pulses supported by the Rydberg blockade mechanism~\cite{Saffman_2016, Levine_2019, Henriet_2020, Madjarov_2020, Cohen_2021, Bluvstein_2021}. We showed how these resources naturally match the requirements to simulate the Trotterized real-time dynamics of general non-abelian gauge fields, thanks to a local encoding where each gauge field acting on a large Hilbert space is simulated using a single atomic qudit. Moreover, we showed how this local mapping allows simulating considerably larger times by reducing the corresponding circuit depths compared to standard qubit-based approaches.

In present work, we extend this quantum simulation architecture to account for matter fields and gauge-matter interactions. This allows us to efficiently simulate general LGTs and to investigate open problems in particle physics, in particular those at finite fermionic chemical potentials. Following a similar hardware-efficient approach as in Ref.~\cite{Gonzalez_2022}, we locally encode and simulate fermionic fields using fermionic atoms, guaranteeing fermionic statistics at the hardware level, leading to a considerable saving in resources compared to non-local qubit encodings. In particular, we extend the {\em fermionic quantum processor} introduced in Ref.~\cite{Gonzalez_2023} and present a full {\em fermion-qudit architecture}. Here, qudits are encoded in fermionic atoms, such that not only the internal but also the motional state of the atom is controlled to process quantum information. Specifically, we implement fermionic tunneling gates through a quantum \textsc{shuttle} based on state-dependent optical tweezers~\cite{Daley_2008}, designed specifically to match the Rydberg-based architecture developed in Ref.~\cite{Gonzalez_2022}, where entangling operations between matter and gauge fields are adapted to fit the \textsc{shuttle} requirements. The proposed local mapping between the simulating and simulated degrees of freedom allows to construct quantum simulation circuits that preserve the locality of the gauge-invariant Hamiltonians, leading to shorter circuit depths and lower simulation errors.

The paper is organized as follows. In Sec.~\ref{sec:general_algorithm}, we summarize the qudit and fermionic gates required to simulate the Trotterized real-time evolution of general LGTs, including both gauge and matter fields, from Higgs bosons to fermions. In Section~\ref{sec:rydberg_quantum_hardware}, we showcase the complete design of the Rydberg-based structure, explaining how the gauge and matter fields can be encoded using trapped atoms in optical tweezers, and how the associated gates are executed with efficiency. We then proceed to demonstrate the abilities of this fermion-qudit processor through two paradigmatic examples. In Sec.~\ref{sec:scalar_QED}, we investigate a minimal realization of scalar quantum electrodynamics through the Abelian-Higgs model, where Higgs fields are coupled to U($1$) gauge fields, one of the simplest models of dynamical matter interacting via gauge fields.
We use our protocol to investigate the dynamics of confinement and string breaking, estimating the required quantum simulation resources for realistic experimental parameters. In Sec~\ref{sec:non_abelian_hadrons}, we show how to prepare non-abelian hadronic states, bound states of a SU($2$) LGT coupled to fermionic matter, using both adiabatic as well as variational techniques. Finally, we utilize our method to study their internal composition via the hadronic tensor, which necessitates calculating time-ordered correlation functions. This operation is difficult to perform classically and highlights the potential of our fermion-qudit processor in computing quantities that have importance in particle physics experiments.

\section{Fermion-qudit gates for digitized gauge theories}
\label{sec:general_algorithm}

In this section, we cover the topic of lattice gauge theories with discrete gauge groups, which can be employed as a digitization of gauge fields based on continuous (Lie) groups. In contrast to previous works that focused on qubit implementations, we identify a native \emph{fermion-qudit} gate set for realizing real-time evolution in discrete Trotter steps. 

We first provide a summary of the hardware requirements in Sec.~\ref{sec:summary_requirements}, which motivates the development of fermion-qudit architecture. Depending on the reader's interest, they may skip the following sections, where we provide detailed derviations of these requirements, and directly proceed to our hardware proposal in Sec.~\ref{sec:rydberg_quantum_hardware}. 
In Sec.~\ref{sec:gauge_fields}, we summarize the protocol introduced in Ref.~\cite{Gonzalez_2022} to simulate non-abelian gauge field dynamics. In Sec.~\ref{sec:higgs_fields} and Sec.~\ref{sec:fermionic_matter_fields}, we extend this approach from pure gauge theories to include matter fields, both bosonic (``Higgs'') and fermionic (``quarks''), respectively.

\subsection{Co-design of Fermi-Qudit Architectures and LGT Algorithms
\label{sec:summary_requirements}}
Non-abelian gauge theories with associated gauge group $G$ coupled to Higgs and/or fermionic matter fields can be simulated efficiently with a fermion-qudit processor. 
Such a processor acts on a hybrid register consisting of qudits $|g_\ell\rangle$ of size $d=|G|$ for every link $\ell$ of a spatial lattice, together with qudits $|g_x\rangle$ of the same size for every site $x$ and/or fermions $|n_{x,\alpha}\rangle$ for the sites.
As we show in the following sections (see Fig.~\ref{fig:lgt_circuit}), the elementary Trotter steps of non-abelian LGTs can be efficiently decomposed into a native set of  basic gates which act either on single degrees of freedom (single links or sites), or pairs of degrees of freedom (link-link, site-site or link-site) of this register. 
Explicitly, we require simple one-body gates $\mathcal{U}$, namely the single-qudit gates  $\mathcal{U}^{(E)}_\ell$ [Eq.~\eqref{eq:group_matrix_elements}],  $\mathcal{U}^{(B)}_\ell$[Eq.~\eqref{eq:group_matrix_elements2}],  $\mathcal{U}^{(M)}_x$ [Eq.~\eqref{eq:Higgs-gates}],  $\mathcal{U}^{(J)}_x$ [Eq.~\eqref{eq:Higgs-gates}] and $\mathcal{U}^{(m)}_x$ [Eq.~\eqref{eq:fermion_mass_term_gate}], as well as a fermionic tunneling gate $\mathcal{U}^{(\text{t})}$ [Eq.~\eqref{eq:tunneling-gate}], which allows us to implement genuine fermionic processes affected by Fermi statistics.
These gates are supplemented by controlled two-qudit gates $C_g\mathcal{U} = \mathcal{U} \otimes \ket{g}\!\bra{g} + \sum_{g^\prime \neq g} \mathbf{1} \otimes \ket{g^\prime}\!\bra{g^\prime}$ [see Eq.~\eqref{eq:controlled-permutation} and~\eqref{eq:controlled-permutation-fermions}], which are used to construct specific entangling operations $\Theta_{\ell|\ell'}$ [Eq.~\eqref{eq:group_multiplication_gate}],  $\Theta_{\ell|x}$[Eq.~\eqref{eq:Higgs-gates}]  and $\mathcal{V}_{x|\ell}$ [Eq.~\eqref{eq:V_x_ell}], whose structure depends on the group $G$.  
In summary, the gate set $\mathcal{G} = \{\mathcal{U},\,C_g\mathcal{U},\, \mathcal{U}^{(\text{t})}\}$
is ideally suited for the digital quantum simulation of non-abelian gauge theories coupled to matter.

We emphasize that the requirements for this approach to simulate finite groups employing qudits are largely independent of the group structure. Given a sufficiently large qudit size, the same algorithms may be used for simulating both abelian and non-abelian gauge theories, making the proposal very versatile. In the Sec.~\ref{sec:rydberg_quantum_hardware}, we show that the whole set $\mathcal{G}$ can be naturally realized on a fermion-qudit architecture using Rydberg atoms in optical traps.

\subsection{Gauge fields}
\label{sec:gauge_fields}

\begin{figure}[ht!]
    \centering
    \includegraphics[width=1.0\linewidth]{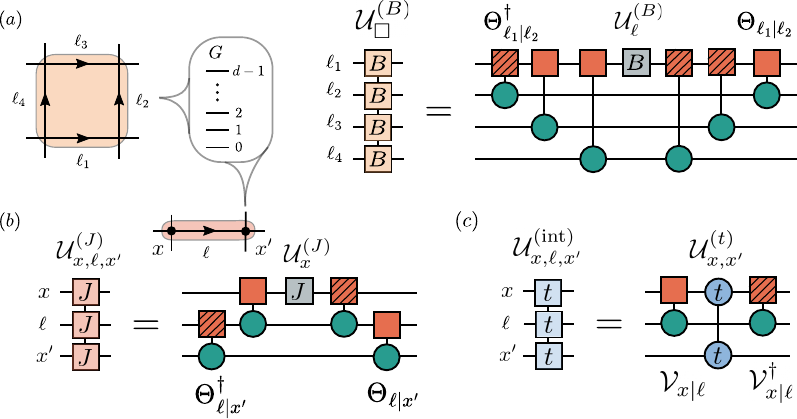}
    \caption{{\bf Trotterized qudit circuits:} (a) For every link $\ell$ of a hypercubic lattice, we consider a $G$-register with $d=|G|$ elements encoded into a single qudit, each of them simulating a local gauge field. A plaquette interaction $\mathcal{U}^{(B)}_\square$ (light orange) is then implemented on four qudits using single-qudit (gray squares) and entangling two-qudit gates $\Theta$ (depicted with green circles and red squares, where the stripes in the former denote the adjoint gate). (b) Similarly, Higgs fields can be digitized and simulated using qudits located on the sites $x$ of the lattice. In this case, the matter-gauge interaction $\mathcal{U}^{(J)}_{x,\ell,x^\prime}$ (light red) is implemented using the same gates as in the pure-gauge case. (c) For SU($N$) fermionic matter fields, we use qudits of size $d=N$. We also require an extra tunneling gate $\mathcal{U}^{(t)}_{x,x^\prime}$ (blue circles) to implement the interaction circuit $\mathcal{U}^{(int)}_{x,\ell,x^\prime}$ (light blue).}\label{fig:lgt_circuit}
\end{figure}

We adopt the  Hamiltonian lattice approach~\cite{kogut1975hamiltonian} for simulating the dynamics of a quantum gauge field theory. Given a gauge group $G$, we consider a general gauge-invariant Kogut-Susskind-type Hamiltonian on a hypercubic lattice with $N_\ell$ links $\ell$, $H_G = \lambda_E H_E + \lambda_B H_B$. Here $H_E$ and $H_B$ denote the ``electric'' and ``magnetic'' contributions, respectively, to the gauge field dynamics, which have the following form
\begin{align}
\label{eq:KS_hamiltonian}
    H_E = \frac{1}{2}\sum_\ell E_\ell^2  \;, && H_B = \sum_\square \left( \mathcal{U}_\square  + \mathcal{U}^\dagger_\square  \right) \;,
\end{align}
where
\begin{align}
    \mathcal{U}_\square = \text{tr} \left[U_{\ell_1} U_{\ell_2} U_{\ell_3}^\dagger U_{\ell_4}^\dagger\right]
\end{align}
acts on the four links $\ell_{1/2/3/4}$ of a plaquette $\square$ [Fig.~\ref{fig:lgt_circuit}(a)], with $\text{tr}\left[..\right]$ denoting a group-dependent trace. Here, $E_\ell$ and $U_\ell$ are operators acting on each link of the lattice. In the non-abelian case, $U_\ell$ corresponds to a matrix of operators (and hence the trace above).

Both the Hilbert space of the theory as well as the precise form of the operators above depend on the structure of the group $G$. In particular, on each link of the lattice, we define a local Hilbert space $\mathcal{H}_G \simeq \mathbb{C}^{|G|}$, whose dimension is given by the size of the group. Here we focus on finite groups of size $|G|=d$, which can be used to approximate continuous groups with an infinite-dimensional Hilbert space. 
Our goal is to identify the common requirements to simulate the time evolution generated by $H_G$ for a general finite group.

We are interested in the evolved state at time $t$, \mbox{$|\psi (t)\rangle = \mathcal{U}_G (t)|\psi(0)\rangle$}, obtained by applying the unitary operator $\mathcal{U}_G (t) = e^{-iH_G t}$ to a given initial state $|\psi(0)\rangle$. At any time, we encode the state on a $G$-register~\cite{Lamm_2019},
\begin{align}
|\psi (t)\rangle = \sum_{\boldsymbol{g}} \psi_t(\boldsymbol{g}) |
\boldsymbol{g}\rangle \;,
\end{align}
where $|\boldsymbol{g}\rangle = \bigotimes_\ell |g_\ell\rangle$ and every $|g\rangle$ denotes the state on each link. The set $\{|g\rangle\}$ forms an orthonormal basis, where each basis state is labeled by a group element $g\in G$.  If we now identify the group basis with a computational basis $\{|j\rangle\}$ ($j=0, \dots, d-1$), we can naturally encode the state $|\psi (t)\rangle$ using $N_\ell$ \emph{qudits} [Fig.~\ref{fig:lgt_circuit}(a)].

The qudit $G$-register is then time evolved by first applying a Trotter decomposition with time step $\delta t$ and error of desired order $\mathcal{O}\left(\delta t^k\right)$ to the time-evolution operator, e.g.~$\mathcal{U}^{(G)}(t) = \left(\mathcal{U}^{(E)}\mathcal{U}^{(B)}\right)^{t/\delta t} + \mathcal{O}(\delta t^2)$ to second order, with $\mathcal{U}^{(E/B)}(\delta t) = e^{-i\lambda_{E/B} H_{E/B} \delta t}$. The locality of the interactions allows us to largely parallelize the latter in terms of the local operations  $\mathcal{U}^{(E)}_\ell(\delta t)= e^{-i\lambda_E E_\ell^2 \delta t/2}$ and $\mathcal{U}^{(B)}_\square(\delta t)= e^{-i\lambda_B \left(\mathcal{U}_\square + \mathcal{U}_\square^\dagger \right) \delta t}$. For any group $G$, the local gates can be written as 
\begin{align}
\label{eq:group_matrix_elements}
\mathcal{U}^{(E)}_{\ell}(\delta t)=
\sum_{h_{\ell}, g_{\ell}\in G}  f^{(E)}(h_{\ell}, g_{\ell},\delta t) \times  |g_{\ell} \rangle_\ell \langle h_{\ell}|
    \;,\\
    \label{eq:group_matrix_elements2}
\mathcal{U}^{(B)}_{\square}(\delta t)=\sum_{g_{\ell_{1,2,3,4}} \in G}
f^{(B)}(g_{\ell_1} g_{\ell_2} g^{-1}_{\ell_3}g^{-1}_{\ell_4},\delta t) \times \\ \nonumber
|g_{\ell_1}, g_{\ell_2},g_{\ell_3}, g_{\ell_4}\rangle \langle g_{\ell_1}, g_{\ell_2}, g_{\ell_3}, g_{\ell_4}| \;,
\end{align}
acting trivially on all other links, where the functions $f^{(E/B)}$ encode the group structure. In the qudit register, $\mathcal{U}^{(E)}_\ell$ corresponds to a general single-qudit gate while $\mathcal{U}^{(B)}_\square$ acts as a diagonal four-qudit gate.

We emphasize that the above structure arises from the defining local symmetry, and hence it is common to all standard (Kogut-Susskind) lattice gauge theories. To be precise, consider a vertex $x$ of the lattice with in- and out-going links $\ell_\text{in}$ and $\ell_\text{out}$ with the associated state $|\{g_{\ell_\text{in}}\}, \{g_{\ell_\text{out}}\} \rangle$. For any $h\in G$, we have a gauge transformation $V_{x,h}$ that acts on that vertex as
\begin{align}
    V_{x,h}|\{g_{\ell_\text{in}}\}, \{g_{\ell_\text{out}}\} \rangle = |\{ g_{\ell_\text{in}}h\}, \{h^{-1}g_{\ell_\text{out}}\} \rangle \;.
\end{align}
The Hamiltonian is gauge-invariant iff $V_{x,h}^\dagger H V_{x,h} = H$ for all $V_{x,h}$. For any group $G$, one can construct local operators $E_\ell^2$, $\mathcal{U}_\square$ that individually remain invariant. In particular, $\mathcal{U}_\square$ depends only on the character $\chi\left(g_{\ell_1} g_{\ell_2} g^{-1}_{\ell_3}g^{-1}_{\ell_4}\right)$ as this combination remains invariant under gauge transformations.

A crucial ingredient in our algorithm is a final decomposition of the four-qudit gate corresponding to the plaquette term~\cite{Zohar_2017a, Zohar_2017b}, which is based on the fact 
that the function $f^{(B)}$ depends only on the product $g_{\ell_1} g_{\ell_2} g^{-1}_{\ell_3}g^{-1}_{\ell_4} \in G$, in particular $f^{(B)}(g) = e^{-2i\lambda_B  \chi(g) \delta t}$, where $\chi$ is the character of the corresponding representation. Since $\mathcal{U}^{(B)}_\square$ is also diagonal in the group basis, it can be realized by first writing the group product into one of the states, then acting with a single-qudit gate $\mathcal{U}^{(B)}_\ell$ that realizes multiplication by $f^{(B)}$ on that qudit, and afterwards undoing the first operation [Fig.~\ref{fig:lgt_circuit}(a)]. The resulting sequence 
\begin{align}
    &|g_{\ell_1}\rangle |g_{\ell_2}\rangle |g_{\ell_3}\rangle |g_{\ell_4}\rangle\nonumber\\ &\;\overset{\Theta}{\rightarrow} |g_{\ell_1} g_{\ell_2} g^{-1}_{\ell_3}g^{-1}_{\ell_4}\rangle |g_{\ell_2}\rangle |g_{\ell_3}\rangle |g_{\ell_4}\rangle \nonumber\\ &\;\overset{\mathcal{U}^{(B)}}{\rightarrow} f^{(B)} (g_{\ell_1} g_{\ell_2} g^{-1}_{\ell_3}g^{-1}_{\ell_4})|g_{\ell_1} g_{\ell_2} g^{-1}_{\ell_3}g^{-1}_{\ell_4}\rangle |g_{\ell_2}\rangle |g_{\ell_3}\rangle |g_{\ell_4}\rangle 
    \nonumber\\ &\;\overset{\Theta^\dagger}{\rightarrow}f^{(B)} (g_{\ell_1} g_{\ell_2} g^{-1}_{\ell_3}g^{-1}_{\ell_4})| g_{\ell_1}\rangle |g_{\ell_2}\rangle |g_{\ell_3}\rangle |g_{\ell_4}\rangle 
\end{align}
shows that
\begin{equation}
\label{eq:plaquette_circuit}
\begin{aligned}
    \mathcal{U}^{(B)}_{\square} = \Theta^\dagger_{\ell_1|\ell_2}  \, \Theta_{\ell_1|\ell_3} \, \Theta_{\ell_1|\ell_4} \, \mathcal{U}^{(B)}_{\ell_1} \, \Theta^\dagger_{\ell_1|\ell_4}  \, \Theta^\dagger_{\ell_1|\ell_3} \, \Theta_{\ell_1|\ell_2} \;.
\end{aligned}
\end{equation}
where we defined elementary two-qudit group multiplication gates acting as $\Theta_{\ell|\ell'}|g_\ell\rangle | g_{\ell'}\rangle = |g_\ell g_{\ell'}\rangle | g_{\ell'}\rangle$. 

Finally, the two-qudit group-multiplication gate can be further decomposed as a product of $d-1$ controlled-permutation gates,
\begin{align}
\label{eq:group_multiplication_gate}
\Theta_{\ell|\ell^\prime} = \sum_{g\in G} \theta_\ell(g) \otimes |g\rangle_{\ell'} \langle g | = \prod_{g\in G} C_{\theta(g)}^{\ell'\rightarrow\ell}(g)\,,
\end{align}
with
\begin{equation}
\label{eq:controlled-permutation}
C_{\theta(g)}^{\ell'\rightarrow\ell}(g) = \theta_{\ell}(g) \otimes \ket{g}_{\ell'}\!\bra{g} + \sum_{g^\prime \neq g} \mathbf{1} \otimes \ket{g^\prime}_{\ell'}\!\bra{g^\prime},
\end{equation}
where $\theta_{\ell}(g)|g_\ell\rangle=|g_\ell g\rangle$ is the right-multiplication group operation.

\subsection{Higgs fields}
\label{sec:higgs_fields}

We now turn our attention to lattice Higgs models. These are theories where gauge fields with gauge group $G$ are interacting with bosonic matter fields. More precisely, we consider Higgs models with a so-called frozen radial mode, where the bosonic fields live at vertices $x$ with an associated Hilbert space spanned by elements $|g_x\rangle$ where $g_x \in G$ belongs to the \emph{same} group as the initial gauge theory [Fig.~\ref{fig:lgt_circuit}(b)]. We then augment the gauge transformation $V_{x,h}$ acting on the state $|g_x, \{g_{\ell_\text{int}}\}, \{g_{\ell_\text{out}}\} \rangle$ around vertex $x$ for an element $h \in  G$ by a contribution from the matter field, namely
\begin{align}
     V_{x,h}|g_x,\! \{g_{\ell_\text{in}}\},\! \{g_{\ell_\text{out}}\} \rangle\! =\! |h^q g_x, \!\{ g_{\ell_\text{in}}h\},\! \{h^{-1}g_{\ell_\text{out}}\} \rangle.
\end{align}
Here, we have chosen that the transformation acts on the left on $|g_x\rangle$, which is sometimes denoted as a Higgs field with gauge group $G_L$ (alternatively, we could consider a right version $G_R$). Additionally, we introduced an integer $q$ that is identified with the charge of the Higgs field, which we set to $q=1$ in the following for simplicity. 

The dynamics of the Higgs fields can be described by the Hamiltonian $H_\text{Higgs} = \lambda_M H_M +\lambda_J H_J$,
\begin{align}
     H_M= \frac{1}{2}\sum_x \Pi^2_x \;, && H_J =  \sum_{\ell=(x,x')} \left( \mathcal{T}_{x\ell,x'} + \mathcal{T}_{x\ell,x'}^\dagger \right) \;,
\end{align}
with the requirement that $H_\text{Higgs}$ is invariant under all combined gauge transformations $V_{x,h}$.
Here, $\Pi^2_x$ is a gauge-invariant term acting on a single vertex $x$, analogous to the $E_\ell^2$ term of the gauge fields, which gives the Higgs fields a mass parametrized by $\lambda_M$. A non-zero value of $\lambda_J$ leads to a gauge-invariant coupling of the gauge field at link $\ell$ to the matter fields at the neighboring sites $x,x'$ via the  operator
\begin{align}
    \mathcal{T}_{x\ell,x'} = \text{tr} \left[\Phi_x^\dagger U_\ell \Phi_{x'} \right] \;,
\end{align}
where $\Phi_x$ are operator-valued matrices for the Higgs fields analogous to the $U_\ell$ for the gauge fields.

It is now straightforward to extend the discussion of the previous section to the implementation of the corresponding new Trotter steps $\mathcal{U}^{(M/J)} = e^{-i\lambda_{M/J}H_{M/J} \delta t}$. Due to the locality of the Hamiltonian, these can again be largely parallelized. For $H_M$, we only need the single-qudit gate $\mathcal{U}^{(M)}_x(\delta t)= e^{-i\lambda_M \Pi_x^2 \delta t}$  acting on the matter qudit associated to site $x$. For $H_J$, we follow a similar strategy as for the plaquette interaction $H_B$ by using the fact that
\begin{align}
    \mathcal{T}_{x\ell,x'} |g_x \rangle |g_\ell \rangle |g_{x^\prime}\rangle = \mathcal{T} (g_x^{-1}g_\ell g_{x^\prime}) |g_x \rangle |g_\ell \rangle |g_{x^\prime}\rangle
\end{align}
is diagonal in our chosen basis, and depends only on the product $g_x^{-1}g_\ell g_{x^\prime}\in G$. Generalizing the group-multiplication gates to also include the matter sites, we thus obtain a decomposition for the three-qudit gate $\mathcal{U}^{(J)}_{x,\ell,x^\prime}(\delta t)= e^{-i\lambda_J ( \mathcal{T}_{x\ell,x'} + \mathcal{T}^\dagger_{x\ell,x'})\delta t}$ into four two-qudit gates $\Theta$ and one single-qudit gate $\mathcal{U}^{(J)}_x$ [Fig.~\ref{fig:lgt_circuit}(b)], namely
\begin{align}
    \mathcal{U}^{(J)}_{x,\ell,x^\prime} = \Theta^\dagger_{\ell|x^\prime}\,\Theta_{x|\ell} \,\mathcal{U}^{(J)}_x\,\Theta^\dagger_{x|\ell}\,\Theta_{\ell|x^\prime} \;,
\end{align}
where $\mathcal{U}^{(J)}_x$ acts just as $\mathcal{U}^{(B)}_\ell$ by replacing $\lambda_B \rightarrow \lambda_J$.

In summary, Higgs fields introduce no genuinely new requirements -- the additional gates
\begin{align}\label{eq:Higgs-gates}
\mathcal{U}^{(M)}_x \;, \quad \mathcal{U}^{(J)}_x \;, \quad \Theta_{\ell|x}
\end{align}
are equivalent to the pure-gauge gates $\mathcal{U}_\ell^{E}, \mathcal{U}_\ell^{B}, \Theta_{\ell|\ell'}$ discussed in the previous section.

\subsection{Fermionic matter fields}
\label{sec:fermionic_matter_fields}

We now further extend our results to include dynamical \emph{fermionic} matter, which is necessary for tackling the relevant theories that form the standard model. 

The total Hamiltonian is now supplemented by two additional terms $H_m$ and $H_\text{int}$
describing a pure matter contribution $(m)$ and the interaction (int) between gauge and matter fields. For brevity and to be explicit, we focus on non-abelian gauge theories with finite group $G \subset {\rm SU}(N)$ coupled to $N$-component staggered fermions in two spatial dimensions, but the generalization to other finite gauge groups and arbitrary dimensions is straightforward. In the following, the fermions are described by annihilation(creation) operators $\psi_{x,\alpha}^{(\dagger)}$ associated with the sites $x = (x_1, x_2)$ of a square lattice and fulfill anti-commutation relations $\{\psi_{x,\alpha}, \psi_{x',\alpha'}^\dagger\} = \delta_{x,x'} \delta_{\alpha,\alpha'}$ with $\alpha = 1, \dots, N$. The staggered fermion mass term is given by
\begin{align}
    H_m = m \sum_{x,\alpha} (-1)^{x_1 + x_2} \, \psi_{x,\alpha}^\dagger \psi^{\vphantom{\dagger}}_{x,\alpha} \;,
\end{align}
with the bare fermion mass $m$. The interaction term acts on the triple $\langle x \ell x'\rangle$ of two neighboring lattice sites $x$ and $x'$ and the connecting link $\ell$ according to
\begin{align}
    H_\text{int} = -t \sum_{\langle x \ell x'\rangle} s_{x x'} \sum_{\alpha,\beta} \left(\psi^\dagger_{x,\alpha} U_{\ell,\alpha\beta} \psi^{\vphantom{\dagger}}_{x',\beta} + \text{h.c.} \right) 
\end{align}
where the sign factor $s_{x x'} = \delta_{x_1,x_1'}(-1)^{x_1} + \delta_{x_2,x_2'}$.
Here $U$ is an operator valued $N\times N$ matrix, which is diagonal in the computational basis of the qudits where it acts multiplicatively according to the fundamental representation of SU($N$), i.e.
\begin{align}
    U_{\alpha\beta} |g \rangle = D_{\alpha \beta}(g) | g \rangle \;,  && g \in G \;,
\end{align}
where $g\in G$ is the group element corresponding to the qudit state $|g\rangle$ and $D(g)$ is the $N\times N$ matrix representing $g$.

In order to simulate real-time evolution 
with a Trotter decomposition, we thus need to implement the additional elementary Trotter steps $\mathcal{U}^{(m)}(\delta t) = e^{-i H_m \delta t}$ and $\mathcal{U}^{(\text{int})}(\delta t) = e^{-i H_\text{int} \delta t}$ for given time step $\delta t$. Let us consider a register of fermions $\{|n_{x,\alpha}\rangle\}$ where the particle-number operator $n_{x,\alpha} = \psi_{x,\alpha}^\dagger \psi_{x,\alpha}$ is diagonal. The contribution of the matter term can then be written as
\begin{align}
     \mathcal{U}^{(m)}  \left[\bigotimes_{x,\alpha}|n_{x,\alpha} \rangle\right] = \bigotimes_{x,\alpha} \left[ e^{-i\delta t \, m(-1)^{x_1 + x_2}  \, n_{x,\alpha}} |n_{x,\alpha} \rangle \right] \;,
\end{align}
i.e. $\mathcal{U}^{(m)} = \prod_{x,\alpha} \mathcal{U}^{(m)}_{x,\alpha}$ decomposes into the relatively simple local single-site phase gates
\begin{align}\label{eq:fermion_mass_term_gate}
\mathcal{U}^{(m)}_{x,\alpha} = e^{-i\delta t \, m(-1)^{x_1 + x_2}  \, \psi_{x,\alpha}^\dagger \psi_{x,\alpha}} \;.
\end{align}
In contrast, $\mathcal{U}^{(\text{int})}$ becomes a more complicated gate involving the fermions on neighboring sites together with the qudit on the link in between.

Following~\cite{Zohar_2017b,Lamm_2019}, we further decompose $\mathcal{U}^{(\text{int})}$ using the unitary operator
\begin{align}
    \mathcal{V}_{x|\ell} = \sum_{g_\ell\in G} |g_\ell \rangle \langle g_\ell| \otimes e^{\sum_{\alpha,\beta} \log [D(g_\ell)]_{\alpha \beta } \psi^\dagger_{x,\alpha} \psi_{x,\beta} } \;,
\end{align}
which acts non-trivially on a single site $x$ and a neighboring link $\ell$. Here the sum runs over all group elements $g_\ell \in G$ (i.e. all computational basis states of the qudit at link $\ell$) and $\log[D]$ denotes the matrix logarithm. The unitarity of $\mathcal{V}$ follows from the unitarity of $D$, which implies the existence of an anti-hermitian logarithm that yields $\left(\log [D(g_\ell)]_{\alpha \beta } \psi^\dagger_{x,\alpha} \psi_{x,\beta}\right)^\dagger = -\log [D(g_\ell)]_{\beta \alpha } \psi^\dagger_{x,\beta} \psi_{x,\alpha}$. The purpose of $\mathcal{V}$ lies in the identities
\begin{align}
    \mathcal{V}_{x|\ell}^\dagger \sum_\alpha\left(\psi_{x,\alpha}^{\dagger} U_{\ell,\alpha \beta} \right) \mathcal{V}_{x|\ell} &= \psi_{x,\beta}^{\dagger} \;, \\ \mathcal{V}_{x|\ell}^\dagger \sum_\beta\left( U^\dagger_{\ell,\alpha \beta} \psi_{x,\beta}  \right) \mathcal{V}_{x|\ell} &=   \psi_{x,\alpha} \;,
\end{align}
which can be proved using the Baker-Campbell-Hausdorff formula. As a consequence, we obtain
\begin{align}
    \mathcal{V}_{x|\ell}^\dagger &\left[ \sum_{\alpha, \beta}\left(\psi^\dagger_{x,\alpha} U_{\ell,\alpha\beta} \psi_{x',\beta} + \text{h.c.} \right) \right] \mathcal{V}_{x|\ell}\nonumber\\ &= \sum_\alpha\psi_{x,\alpha}^\dagger \psi_{x',\alpha} \;,
\end{align}
which reduces $H_\text{int}$ to a non-interacting fermion hopping Hamiltonian. We thus have the decomposition [Fig.~\ref{fig:lgt_circuit}(c)]
\begin{align}
\label{eq:gauge_matter_gate}
    \mathcal{U}^{(\text{int})} &= \prod_{\langle x \ell x'\rangle} \left\{\mathcal{V}_{x|\ell} \, \mathcal{U}^{(\text{t})}_{xx'} \, \mathcal{V}^\dagger_{x|\ell} \right\} \;, \\  \label{eq:tunneling-gate}\mathcal{U}^{(\text{t})}_{xx'} &=  e^{-i\delta t \,  s_{xx'} \sum_\alpha \left(\psi_{x,\alpha}^\dagger \psi_{x', \alpha} +\text{h.c.} \right)} \;.
\end{align}

In summary, the additionally required gates are: the single-site phase gates $\mathcal{U}^{(m)}_{x,\alpha}$, the fermionic tunneling gates $\mathcal{U}^{(\text{t})}_{xx'}$ and the special fermion-qudit entangling gates $\mathcal{V}_{x,\ell}$. We further note that the latter are very similar to the group-multiplication gates whose implementation with qudits we have already discussed for the pure gauge theory~\cite{Gonzalez_2022}. In particular, we can further decompose the gate
\begin{align}\label{eq:V_x_ell}
    \mathcal{V}_{x|\ell} = \prod_{g\in G} C^{\ell\rightarrow x}_{V_x(g)}(g) \;,
\end{align}
where
\begin{equation}
\label{eq:controlled-permutation-fermions}
C^{\ell\rightarrow x}_{V_x(g)}(g) = V_{x}(g) \otimes |{g}\rangle_{\ell'}\langle {g}| + \sum_{g' \neq g} \mathbf{1} \otimes |g' \rangle_{\ell'}\langle g'|,
\end{equation}
are controlled-unitary gates, which are again controlled by the qudits, but this time act with the unitary $V_x(g) =  e^{\sum_{\alpha,\beta} \log [D(g)]_{\alpha \beta } \psi^\dagger_{x,\alpha} \psi_{x,\beta} }$ on the involved fermions.

\section{Rydberg-based fermion-qudit quantum processor}
\label{sec:rydberg_quantum_hardware}

In this section, we describe a quantum hardware architecture based arrays of Rydberg atoms trapped in optical tweezers. It implements the fermion-qudit processor described in Sec.~\ref{sec:general_algorithm}, and is tailored to efficiently run digital quantum simulations of LGTs based on the algorithms described above. 

In Ref.~\cite{Gonzalez_2022}, we developed a qudit gate set that naturally matches the requirements to digitally simulate the dynamics of non-abelian gauge fields, using quantum circuits that preserve the locality of the LGT Hamiltonian. Following a similar hardware-efficient approach,  we extend this architecture by developing the required gates to couple gauge fields to dynamical matter. In Sec.~\ref{sec:rydberg_qudits}, we show how the internal atomic structure of Rydberg atoms allows us to naturally encode the $G$-registers required to simulate finite gauge groups, as well as both Higgs and fermionic matter fields, and to perform single-qudit operations in a combined fermion-qudit register. In Sec.~\ref{sec:fermionic_shuttling}, we extend the \textsc{shuttle} protocol introduced in Ref.~\cite{Gonzalez_2023} to implement fermionic tunneling gates. Finally, in Sec.~\ref{sec:rydberg_gates} we complete the gate set required to Trotterized the dynamics of general LGTs by introducing entangling operations between gauge and matter fields.

\subsection{Fermion-qudit atomic register}
\label{sec:rydberg_qudits}

\subsubsection{Local field encoding}

In the previous section, we discussed the Hilbert space structure of general LGTs. In particular, for a general finite group $G$ of size $|G| = d$, the pure-gauge part of the LGT Hamiltonian~\eqref{eq:KS_hamiltonian} acts on $\mathcal{H}=\bigotimes_{\ell} \mathbb{C}^d$, where the tensor product is taken over the links $\ell$ of a hypercubic lattice whose vertices can be labeled as elements of $\mathbb{Z}^D$, where $D$ is the spatial dimension of the lattice. In order to efficiently simulate LGTs in a scalable manner, using quantum hardware that possesses the same Hilbert space structure offers a clear advantage. Rydberg atoms trapped in optical tweezers provide a natural choice, thanks to the possibility of spatially arranging the atoms into any desired geometry. Moreover, every atom possesses an internal $\mathcal{H} = \mathbb{C}^d$ Hilbert space associated with its long-lived hyperfine ground-state manifold where the $G$-register qudit can be encoded while entangling gates can be designed by turning on and off strong interactions of the excited Rydberg states.

\begin{figure}[ht!]
    \centering
    \includegraphics[width=0.9\linewidth]{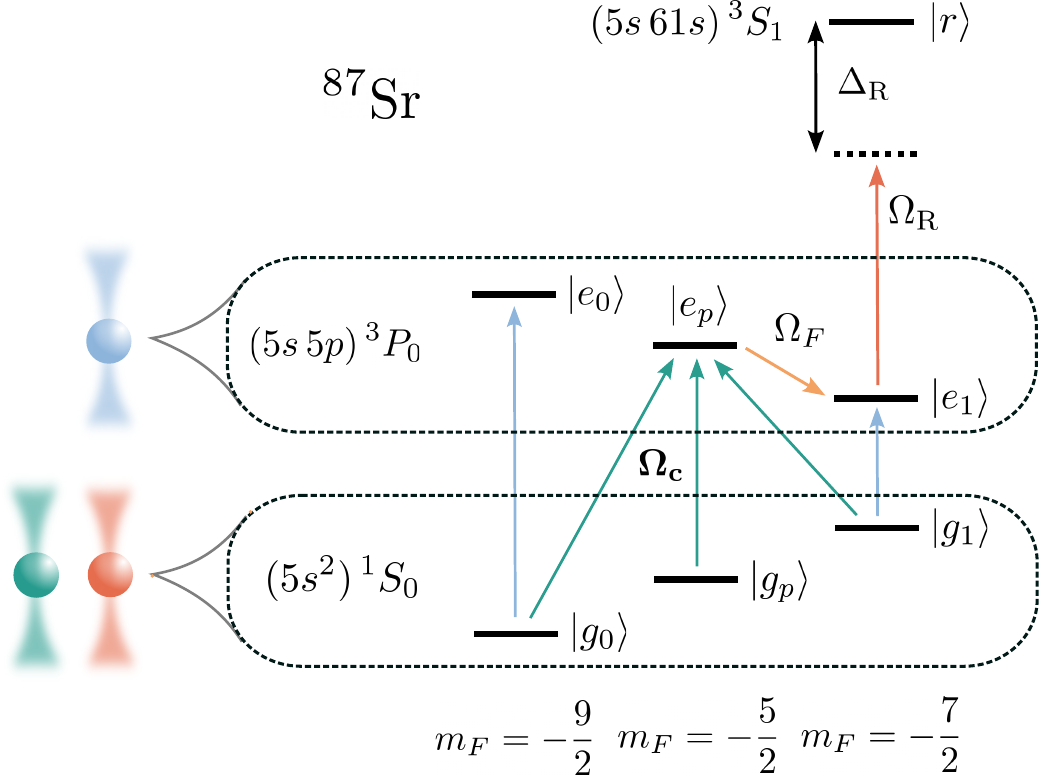}
    \caption{{\bf Atomic structure for ${}^{87}$Sr:} both gauge and matter field basis states can be encoded in the ground-state hyperfine manifold of ${}^{87}$Sr, $\{|g_i\rangle\}_{i=0,...,d-1}$, corresponding to the electronic configuration $(5s^2)\,{}^1 S_0$ and exemplified in the figure for the case $d=2$, where atoms are trapped using storage tweezers (red and green denote (fermionic) matter and gauge fields, respectively). The latter is coupled to the meta-stable excited manifold $(5s5p)\,{}^3 P_0$, where $\mathbf{\Omega_c}$ denotes the vector of the corresponding Rabi couplings, trapped in this case by transport tweezers (blue). This excited manifold $\{|e_i\rangle\}_{i=0,...,d-1}$ serves different purposes. First, holonomic operations are implemented in the ground-state qudit using the auxiliary states $|e\rangle_p$ and $|g\rangle_p$. Second, in the fermionic case, atoms can be moved in the transport tweezer. Finally, entangling operations are implemented by passing through the state $|e\rangle_{d-1}$, dressed by the Rydberg state $|r\rangle$ in the $(5s61s)\,{}^3 S_1$ manifold. The former is coupled to $|e_p\rangle$ using, e.g., a radio-frequency field with Rabi coupling $\Omega_F$, and Rydberg-dressed by coupling to $|r\rangle$ with detuning $\Delta_{\rm R}$ and Rabi frequency $\Omega_{\rm R}$.
    }
    \label{fig:atomic_structure}
\end{figure}

In Fig.~\ref{fig:atomic_structure}, we present a possible atomic level scheme using $^{87}$Sr as an example. This includes auxiliary levels used to implement single-qudit and entangling gates, to be discussed below. Strontium as an alkaline-earth atom possesses a nuclear spin of $I=9/2$, allowing to locally encode for instance $\mathbb{Z}_8$ or $Q_8$ gauge fields, the latter corresponding to the minimal digitization of the SU($2$) LGT with $d = 8$ elements~\cite{Gonzalez_2022}. As  recently demonstrated, Sr atoms can be trapped in their ground state manifold $^1S_0$ using optical tweezers~\cite{Cooper_2018}, which we will call in the following \textit{storage} tweezers, which can be then rearranged to the desired geometry and allow for single-site resolution~\cite{Covey_2019}. As an alternative to the storage tweezers, one could also use optical lattices for storage with the benefit of gaining stability, as we discuss below.

Similar to gauge fields, local matter fields can also be encoded into single atoms, placed in this case on the $N_s$ sites of a $D$-dimensional hypercubic lattice. As described in the previous section, Higgs fields are digitized in the same manner as the gauge fields they are couple to, and the whole gauge-invariant state can be encoded using $(D+1)N_s$ atoms. For the case e.g. $D = 2$, they form a homogeneous Lieb lattice composed of $3N_s$ qudits.

Fermionic matter fields with $N$ components can similarly be encoded using qudits of size $N$. Driven by a hardware-efficient use of resources, we consider fermionic atoms trapped in optical tweezers, such as $^{87}$Sr, providing a local mapping to the simulated degrees of freedom and reducing the corresponding quantum simulation circuit depths, as we will show below. Fig.~\ref{fig:atomic_structure} shows the required atomic structure to store and manipulate SU($2$) fermionic fields, with $N = 2$. Finally, note that although we have assumed for simplicity that both matter and gauge fields are encoded using the same type of atoms, gauge fields can be encoded also using bosonic atoms, leading to arrays of atomic mixtures~\cite{Stingh_2022}.

\subsubsection{Single-qudit gates}

In Ref.~\cite{Gonzalez_2022} we introduced a protocol to implement general single-qudit operations that can be applied both to gauge and matter qudits, accounting for the required single-qudit gates listed in the previous section:  $\mathcal{U}^{(B)}_\ell, \mathcal{U}^{(E)}_\ell, \mathcal{U}^{(M)}_x, \mathcal{U}^{(J)}_x$ and $\mathcal{U}^{(m)}_x$. 
This can be done by first decomposing the corresponding unitary as a product of $d(d-1)/2$ operations acting non-trivially only between two consecutive levels, 
this is, $U = \prod_{k}\tilde{U}_k$ $\in$ SU($d$), with $\tilde{U}_k = \mathbf{1}_{j_k - 1}\oplus U_k \oplus \mathbf{1}_{d - j_k - 1}$, where $U_k \in {\text{SU}}(2)$ acts on a two-dimensional subspace $\mathcal{H}_j$ spanned by $\{\ket{g_{j_k}}, \ket{g_{j_k + 1}}\}$. 
Each SU(2) operation $U_k$ can be implemented holonomically with the help of auxiliary states, denoted $\ket{g_p}$ and $\ket{e_p}$ in Fig.~\ref{fig:atomic_structure}, using pairs of laser pulses characterized by Rabi frequencies $\mathbf{\Omega_c}^{(j_k)} = (\Omega^{(j_k)}_{0}, \Omega^{(j_k)}_{1}, \Omega^{(j_k)}_{p})$, which defines a parameter manifold $\mathcal{M}$ for the Hamiltonian~\cite{Gonzalez_2022}. 

Let us consider in particular the following single-particle Hamiltonian,
\begin{equation}
\begin{aligned}
H(\mathbf{\Omega_c}) = \frac{1}{2}\ket{e_{p}}&\Big(\Omega_{0}\bra{g_0} + \Omega_{1} \bra{g_1}\\
&+ \Omega_{p} \bra{g_p} \Big) + \text{H.c.},
\end{aligned}
\end{equation}
where we dropped the index $j_k$ to simplify the notation. For every parameter $\mathbf{\Omega}_c \in \mathcal{M}$, $H(\mathbf{\Omega_c})$ possesses four eigenstates: two dark states $\ket{\psi_{a=0,1}(\mathbf{\Omega})}$ at zero energy and two states with energies $\pm \Delta(\mathbf{\Omega_c}) = \pm \sqrt{|\mathbf{\Omega_c}|^2}$, with a gap that will remain open as long as $\Omega_p \neq 0$. ${\rm SU}(2)$ operations $u:\,\mathcal{H}_j \to \mathcal{H}_j$ can then be implemented via closed loops $\gamma_C: [t_0, t_1] \to \mathcal{M}$, with $\gamma(t_0) = \gamma(t_1)$. More specifically, if the parameters in the Hamiltonian change adiabatically along the loop, $u$ can be written as~\cite{Zanardi_1999}
\begin{equation}
u = {\rm \mathbf{P}}\, {\rm exp} \left(-i\int_{\gamma_C}\sum_\mu A^\mu {\rm d}\Omega_\mu\right),   
\end{equation}
where $(A^\mu)_{ab} = \bra{\psi_a(\mathbf{\Omega})}\partial/\partial \Omega_\mu\ket{\psi_b(\mathbf{\Omega})}$ is the corresponding connection.

One can use in particular Gaussian pulses for $\Omega_{a = 0, 1}$ that overlap in time and have the form
\begin{equation}
    \label{eq:gaussian_pulses}
    \Omega_{a}(t) = \Omega \, \text{e}^{-(t - \tau_{a})^2/(T/10)^2} \text{e}^{-i\varphi_{a}},
\end{equation} 
where $T$ is the pulse time window, giving rise to the following gates,
\begin{equation}
\label{eq:u_delta_gamma}
    u(\alpha,\delta)=
    \begin{pmatrix}
      \cos\alpha & e^{-i\delta}\sin\alpha\\
      -e^{i\delta}\sin\alpha & \cos\alpha
    \end{pmatrix}\,.
\end{equation}
Here, $\delta = \varphi_{1} - \varphi_{0}$ is the phase difference between the pulses and $\alpha$ can be tuned between $-\pi / 2$ and $\pi / 2$ by adjusting the time overlap $\tau = \tau_{1} - \tau_{0}$. For $\delta = \pi/2$ ($\pi$) we obtain in particular rotations around the $X$ ($Y$) axis. In summary, any single-qudit operation can be implemented using at most $3d(d-1)/2$ pairs of pulses, since any ${\rm SU}(2)$ operation can be written in terms of at most three rotations. The total time required to implement a general single-qudit gate is thus $\mathcal{O}(d^2)T$, and we impose $\Omega_c T \gg 1$, enforcing the adiabaticity required for the holonomic implementation and thereby supressing the associated errors.

\subsection{Fermionic \textsc{shuttle} gates}
\label{sec:fermionic_shuttling}

\begin{figure}[ht!]
    \centering
    \includegraphics[width=1.0\linewidth]{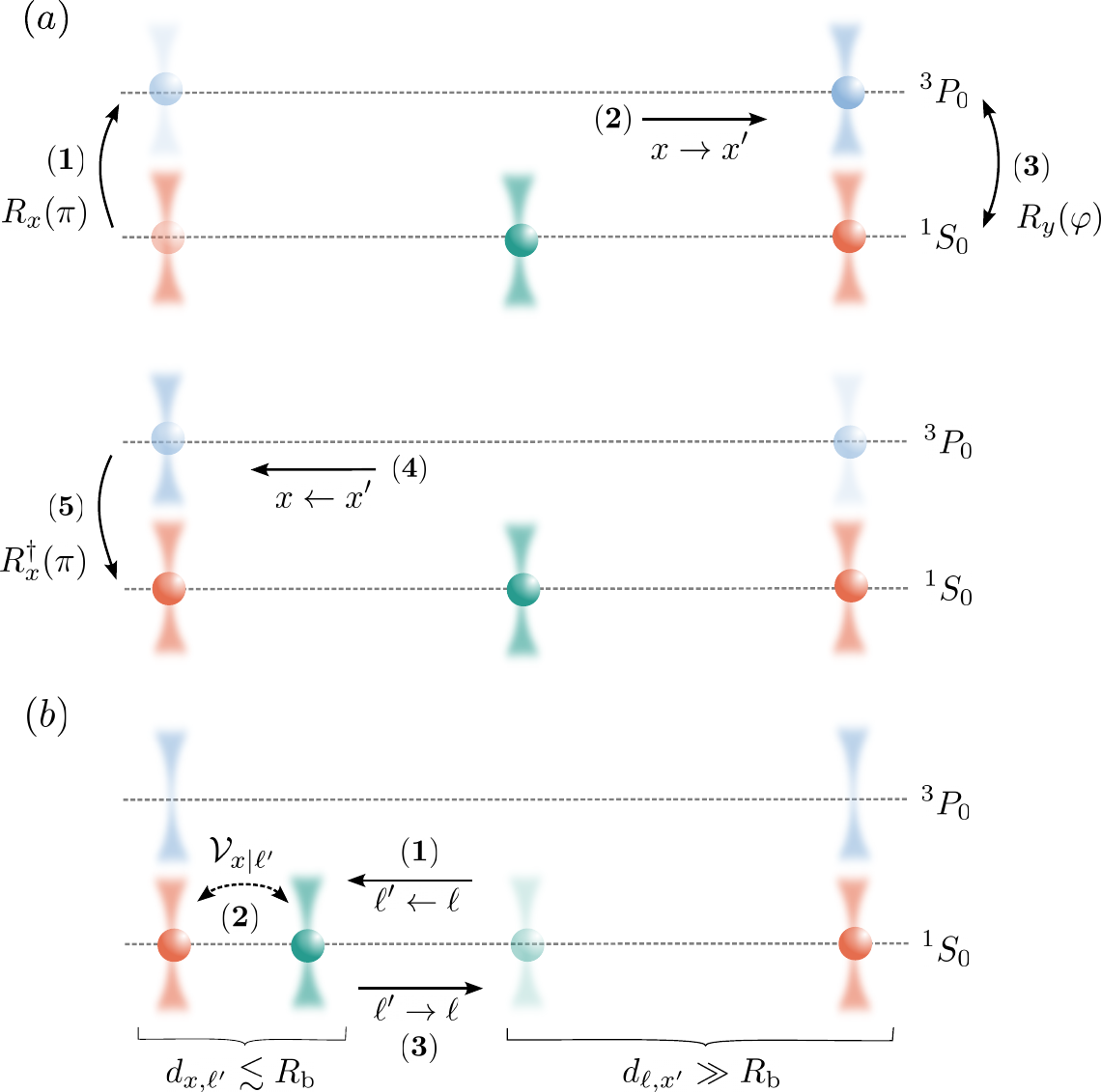}
    \caption{{\bf Gauge-matter interactions:} ${}^{87}$Sr atoms in the ground-state manifold ${}^{1}S_0$, encoding both fermionic (red) and gauge fields (green), are trapped using optical tweezers and arranged in the desired geometry, far enough to avoid any cross talk when excited to Rydberg states. (a) The fermionic \textsc{shuttle} is implemented by {\bf (1)} first transferring the atoms on the sites to the transport tweezers (blue) using a rotation $R_x(\pi)$ to the meta-stable manifold ${}^{1}P_3$. {\bf (2)} The transport tweezers are then displaced in one direction ($x\rightarrow x^\prime$) and {\bf (3)} the atoms rotated with $R_y(\varphi)$ between the the storage and the transport tweezers. Finally, {\bf (4)} the transport tweezers are displaced back ($x\rightarrow x^\prime$) and {\bf (5)} cleaned using $R^\dagger_x(\pi)$. (b) The entangling gate $\mathcal{V}_{x|\ell}$ is obtained by {\bf (1)} first moving the tweezers ($\ell^\prime\leftarrow\ell$) to bring the atoms within the blockade radius, {\bf (2)} then applying the corresponding sequence of pulses (see main text) and finally {\bf (3)} bringing the atoms back ($\ell^\prime\rightarrow\ell$). The complete gauge-matter interaction gate $U^{(\rm int)}$ Eq.~\eqref{eq:gauge_matter_gate} is obtained by applying the protocol in (b), then (a) and finally (b) again but using $\mathcal{V}^\dagger_{x|\ell}$ instead.
    }
    \label{fig:gauge_matter_gate}
\end{figure}

Using fermionic atoms to encode fermionic matter fields allows to implement the tunneling gate in Eq.~\eqref{eq:gauge_matter_gate} in a natural and hardware-efficient manner. To be concrete, we continue with the fermionic isotope ${}^{87}\text{Sr}$ as an example. 
To implement fermion tunneling, we follow the scheme developed in Ref.~\cite{Gonzalez_2023} and consider two independent tweezers: the storage tweezer traps the ground-state manifold ${}^1S_0$ as mentioned above, and an additional \textit{transport} tweezer traps the meta-stable excited manifold ${}^3P_0$ (Fig.~\ref{fig:atomic_structure}). 
Since we need to correctly control the motional degrees of freedom to properly simulate fermionic statistics, fermionic atoms trapped in the storage tweezers have to be first cooled down to their motional ground state, as it was recently demonstrated~\cite{Spar_2021, Yan_2022}. The desired tunneling in a given direction can now be implemented by shining a series of pulses, interlaced with spatial movements of the transport tweezers [Fig.~\ref{fig:gauge_matter_gate}(a)].

Let us make this more explicit for two lattice sites $x,x'$ and four fermion modes $\psi_{x,\alpha}$, $\psi_{x^\prime,\alpha}$, $\tilde{\psi}_{x,\alpha}$, $\tilde{\psi}_{x^\prime,\alpha}$, where $\psi_\alpha$ $(\tilde{\psi}_\alpha)$ corresponds to the ${}^{1}S_0$ $({}^{3}P_0)$ manifold for an hyperfine state $\ket{g_\alpha}$ ($\ket{e_\alpha}$) encoding a given fermionic mode $\alpha\in\{1,\dots, N\}$ (Fig.~\ref{fig:atomic_structure}). In this setup, {\bf (1)} we first perform a $\pi$-pulse rotation on site $x$, i.e. $R_x(\pi) = e^{-i(\pi/2)(\psi_{x,\alpha}^\dagger \tilde{\psi}_{x,\alpha}^{\vphantom{\dagger}} + \text{H.c.})}$, {\bf (2)} we then move the transport tweezer such that $\tilde{\psi}_{x,\alpha} \rightarrow \tilde{\psi}_{x^\prime,\alpha}$, denoted $T_{x\rightarrow x^\prime}$. {\bf (3)} Then we perform a second pulse, this time on site $x'$, for a desired rotation parametrized by $\varphi$, namely $R_y(\varphi) = e^{-\varphi/2\left(\psi_{x^\prime,\alpha}^\dagger \tilde{\psi}_{x^\prime,\alpha}^{\vphantom{\dagger}} - \text{H.c.}\right)}$, {\bf (4)} after which we bring the transport tweezer back into its original position with $T_{x^\prime\rightarrow x}$, {\bf (5)} and finally undo the initial $\pi$-pulse. Putting it all together, this protocol realizes
\begin{align}
    &e^{-i\varphi/2 \left(\psi_{x,\alpha}^\dagger \psi_{x', \alpha} +\text{H.c.} \right)} \nonumber\\ &\qquad= R^\dagger_x(\pi) \, T_{x^\prime \rightarrow x} \, R_y(\varphi) \, T_{x\rightarrow x^\prime} \, R_x(\pi),
\end{align}
Note that this operation can be performed simultaneously for every $\alpha$, realizing thus $\mathcal{U}^{\rm (t)}_{xx^\prime}$. Moreover, this can be parallelized on pairs of lattice sites $(x,x')$ due to the possibility of dynamically arranging tweezer arrays loaded with atoms in a coherent manner~\cite{Bluvstein_2021}. 

Fermionic tunneling could be alternatively implemented using storage tweezers only, by bringing them first close enough such that fermions can tunnel between different tweezers. As we mentioned above, however, it is crucial that fermions remain at all times in the motional ground state of the trap, requiring deep trapping conditions. 
In principle, this is compatible with direct tunneling between tweezers~\cite{Murmann_2015, Bergschneider_2019, Becher_2020, Spar_2021, Young_2022, Yan_2022}, leading the alternative \textsc{merge} gate discussed in~\cite{Gonzalez_2023}. While we focus on realizations based on tweezer arrays here, we note that one can realize similar fermionic operations, including entangling gates, with alternative setups involving optical lattices, which were orginally developed in the context of bosonic atoms \cite{Jaksch_1999, Mandel_2003_0, Mandel_2003, Daley_2008, belmechri2013microwave, Robens_2015, Lam_2021, Zhang_2022}. More generally, one could combine the advantages of tweezer array with optical lattices~\cite{private_Bloch}. The latter have the advantage of guaranteeing the same potential depth for each minima, minimizing the possibility of dephasing that could arise between different tweezers due to relative fluctuations in their intensity. One could then use optical lattices to store the atoms, serving as a fermionic register, which can be the transferred to optical tweezers to implement the required gates, leveraging their flexibility to bring any pair of atoms together to implement non-local gates in a parallel fashion.

\subsection{Rydberg blockade and entangling gates}
\label{sec:rydberg_gates}

We finish this section by summarizing the protocol introduced in Ref.~\cite{Gonzalez_2022} to implement controlled-unitary two-qudit gates in the Rydberg architecture, required to construct the entangling gates $\Theta_{\ell|\ell'}$ and $\Theta_{\ell|x}$. This protocol, based on the Rydberg blockade mechanism, can be easily adapted to implement the fermion-gauge entangling gate $\mathcal{V}_{x|\ell}$, as we show below. To see this, consider first the interactions between a pair of atoms excited to the Rydberg states $\ket{r}_x$ and $\ket{r}_{x^\prime}$, where $x$ and $x^\prime$ can be either sites or links, given by
\begin{equation}
H^{(\ell, \ell^\prime)}_\text{int} = \frac{V_0}{d_{x,x^\prime}^6} \left(\ket{r}\bra{r}\right)_x \otimes \left(\ket{r}\bra{r}\right)_{x^\prime},
\end{equation}
where the interaction strength decays as a power law with the distance $d_{x,x^\prime}$ between the atoms. In order to implement a controlled two-qudit gate, the tweezer geometry is first rearranged to bring the corresponding pair of atoms within each other's blockade radius~\cite{Bluvstein_2021}, with $V\equiv V_0 / d^6_{i,j}\gg \Omega$, such that only one of them can be excited to the Rydberg manifold [Fig.~\ref{fig:gauge_matter_gate}(b)]. We assume that they are far enough from the other atoms in the array, such that any cross talk can be safely neglected. 

To implement any controlled unitary operation $C_g\mathcal{U}$ we perform the following steps~\cite{Gonzalez_2022}: $(i)$ realize the single-qudit operation $U$ on the target atom at site $x$, as described above. $(ii)$ excite the qudit $x^\prime$ from the control state $|g\rangle$ to a Rydberg state $\ket{r}_{x^\prime}$. $(iii)$ perform $U^\dagger$ on the fermionic atom, this time passing through the Rydberg state $\ket{r}_x$. $(iv)$ Bring the qudit $x$ from the Rydberg state to $|g\rangle$. This protocol implements $U$ on the target atom \textit{if and only if} the control atom is in the state $\ket{g}$ since, in any other case, the Rydberg blockade mechanism prevents the second step of the protocol. The group-multiplication gates $\Theta_{\ell|\ell'}$ and $\Theta_{\ell|x}$ are then obtained by repeating the protocol above for every value of $g$, with $U = \theta_g$ (Eq.~\eqref{eq:group_multiplication_gate}).

If the target atom encodes a fermionic field, the protocol has to be slightly modified to make it compatible with the fermionic \textsc{shuttle}. As we mentioned above, it is crucial that fermions remain in their motional ground state within the trap. Exciting a fermionic atom to a Rydberg state, where it is not trapped anymore, could therefore introduce decoherence in the motional degrees of freedom when brought back to the ground state. For this reason, in the above protocol, instead of coupling the target fermionic atom directly to a Rydberg state~\cite{Madjarov_2020}, one possible alternative is to couple it to a dressed Rydberg state~\cite{Henkel_2010}. To illustrate this possibility, we consider an additional laser that dresses one state of the ${}^3P_0$ manifold, where is trapped in the transport tweezer, with a Rydberg state (see Fig.~\ref{fig:atomic_structure}) and bring the relevant qudit $\ell$ close to the site $x$ such that the (reduced) Rydberg interaction with the dressed ${}^3P_0$ state is strong enough to create a blockade. The protocol to implement $C_gU$ then goes goes as described above with the only difference that $U^\dagger$ on step $(iii)$ is implemented by passing through the dressed state. Note that, thanks to the constant dressed Rydberg potential within the blockade radius~\cite{Henkel_2010}, the atoms experience no force and remain in their motional ground states. Moreover, since all the auxiliary states required to perform single-qudit operations are also trapped by either the storage or the transport tweezer (Fig.~\ref{fig:atomic_structure}), the atom is never free, keeping the motional state under control. We leave a detailed comparison of the direct Rydberg protocol and the dressed version outlined here for future work.

The gate $\mathcal{V}_{x|\ell}$ is then obtained by repeating this protocol for every value of $g$ and using the corresponding unitaries on the fermions. After having entangled the gauge field at $\ell$ with the fermions at $x$, we use the shuttling protocol to implement the tunneling gate. Finally, we undo the entangling qudit-fermion gate $\mathcal{V}^\dagger_{x|\ell}$ by an analogous protocol, obtaining the complete gauge-matter interaction gate $\mathcal{U}^{\text{(int)}}$~\eqref{eq:gauge_matter_gate}.

As a final comment, we note that the different fermionic species $\alpha$ trapped in the same tweezer interact locally, leading in principle to unwanted phase shifts. This shift is particularly simple in the case of alkaline-earth atoms such as $^{87}$Sr, which possesses a SU($N$) symmetry in its ground state such that all interactions between every pair of hyperfine states $\alpha$ are controlled by a single scattering length~\cite{Zhang_2014, Goban_2018}. For given time $t$, it is thus determined by $e^{-i\omega t \sum_{\alpha \beta} n_\alpha n_\beta}$ with occupation $n_\alpha = \psi^\dagger_\alpha \psi_\alpha$ and a single frequency $\omega$ independent of $\alpha,\beta$. Above, we have neglected this error as a first approximation. Ideally, since these phase shift errors are coherent processes, we can correct for them by applying entangling gates $\prod_{\alpha \beta} e^{-i\theta n_\alpha n_\beta}$ with $\theta = -\omega t$ that cancel them using e.g. dressed-Rydberg interactions between different fermionic species as explained above.

\section{Scalar quantum electrodynamics}
\label{sec:scalar_QED}

In this section, we discuss in detail the digital quantum simulation of a U($1$) LGT, describing electromagnetic interactions in the Standard Model of particle physics, coupled to a scalar matter field. In Sec.~\ref{sec:U(1)_lgt}, we specified the requirements of our general qudit protocol for this particular case, where the U($1$) dynamics are approximated by its $\mathbb{Z}_d$ subgroups. In Sec.~\ref{sec:AH_model} we show how, in the abelian case, Higgs fields can be easily integrated out and how the corresponding matter-gauge coupled dynamics can be quantum simulated using the same resources as for the pure-gauge case.  In Sec.~\ref{sec:physics_string_breaking}, we analyze a minimal example of the model and show how string-breaking dynamics could be observed using our protocol, and how the continuous U($1$) dynamics are recovered for relatively small values of $d$.

\subsection{U($1$) lattice gauge theory}
\label{sec:U(1)_lgt}

As our first example, consider the gauge group U($1$). Formally, we have the group element states $|\varphi\rangle$ with $\varphi \in [-\pi,\pi]$ labelling all elements  $e^{i\varphi}\in$ U($1$). The pure gauge Hamiltonian is determined by the electric operator $E$ and the link operator $U$, which act as
\begin{align}
    E|\varphi\rangle = -i\partial_\varphi |\varphi\rangle  \;, && U|\varphi\rangle = e^{i\varphi} |\varphi\rangle
\end{align}
and fulfill the algebra $[E,U] = U$.
We also have a dual basis spanned by the ``representation'' states $|n\rangle = 1/(2\pi)\int \! {\rm d} \varphi \,e^{i\varphi n} |\varphi\rangle$ with $n\in \mathbb{Z}$, where $E$ becomes diagonal and $U$ acts as a raising operator,
\begin{align}
    E|n\rangle = n | n\rangle \;, && U|n\rangle = |n+1 \rangle \;.
\end{align}

We digitize U($1$) via its subgroups $\mathbb{Z}_d$ consisting of the $d$ elements $e^{-2\pi i n/d}$ with $n=0, \dots, d-1$. By slight abuse of notation, we denote the corresponding group state also by $|\varphi\rangle$, omitting the $d$-dependence. Again we have a dual basis spanned by the analogous states $|n\rangle =(1/d)\sum_\varphi e^{i\varphi n} | \varphi\rangle$. The resulting digitization of $U$ is conventionally denoted by $Q$ with 
\begin{align}
    Q|\varphi \rangle = e^{i\varphi}|\varphi \rangle \;, && Q |n \rangle = |(n+1) \mod d\rangle \;.
\end{align}
The digitization of $E$ is less obvious since it acts as a derivative in group space. While the situation for general non-abelian groups is more involved, there is a simple work-around for U($1$) by considering the limit 
\begin{align}
    \frac{d^2}{(2\pi)^2} \left[1-\cos(2\pi E/d)\right] \rightarrow \frac{1}{2} E^2 && \text{for} && d\rightarrow\infty \;.
\end{align}
This shows that the modified Hamiltonian 
\begin{align}
    H_{\mathbb{Z}_d} &= -\frac{\lambda_E d^2}{2\pi^2} \sum_{\ell}\left(P_\ell + P_\ell
^\dagger\right) \nonumber \\ &\quad+ \lambda_B \sum_\square \left(Q_{\ell_1} Q_{\ell_2} Q_{\ell_3}^\dagger Q_{\ell_4}^\dagger + \text{h.c.}\right)
\end{align}
with operators $P$ defined by
\begin{align}
    P|n\rangle = e^{2\pi i n/d}|n\rangle \;, && P|\varphi \rangle = |\varphi + \frac{2\pi}{d}\rangle\;,
\end{align}
generates the same physics as the U($1$) theory in the limit $d \rightarrow \infty$.

Turning to our qudit approach, we represent the gauge field state on every link by a qudit of size $d$. The required local gates are now explicitly given by
\begin{align}
f_E(\varphi', \varphi) &= \langle \varphi'|e^{+i\lambda_E d^2 / (2\pi^2) (P + P^\dagger)\delta t}|\varphi\rangle \;,\\
f_B(\varphi) &= \langle \varphi|e^{-i\lambda_B (Q+Q^\dagger)\delta t}|\varphi\rangle  = e^{-2i\lambda_B \cos(\varphi)\delta t}  \label{eq:fB_Zd}\;.
\end{align}
The above discussion further shows that the electric single-qudit gate is diagonal in Fourier space, i.e.
\begin{align}
    f_E(n',n) &= \frac{1}{d}\sum_{\varphi',\varphi} e^{i\left(\varphi'n'- \varphi n\right)} f_E(\varphi', \varphi) \nonumber \\&= \delta_{n',n} e^{2i\lambda_E d^2/(2\pi^2) \cos(2\pi n/d)\delta t}  \label{eq:fE_Zd}\;.
\end{align}
Consequently, two types of single-qudit gates, namely the diagonal phase gates specified by Eqs.~\eqref{eq:fE_Zd} and~\eqref{eq:fB_Zd} together with a Fourier transform gate, are sufficient for our purposes. Following the general strategy, these single-qudit gates need to be supplemented by a two-qudit gate that realizes group multiplication. For the case of $\mathbb{Z}_d$, this is given by a controlled-addition (CADD) operation, i.e.
\begin{align}
\label{eq:CADD}
    \Theta(\ell | \ell')|\varphi_\ell\rangle |\varphi_{\ell'}\rangle = | (\varphi_{\ell} + \varphi_{\ell'}) \mod 2\pi \rangle|\varphi_{\ell'} \rangle  \;.
\end{align}

\subsection{Abelian-Higgs model}
\label{sec:AH_model}

\begin{figure}
    \centering
    \includegraphics[width=1.0\linewidth]{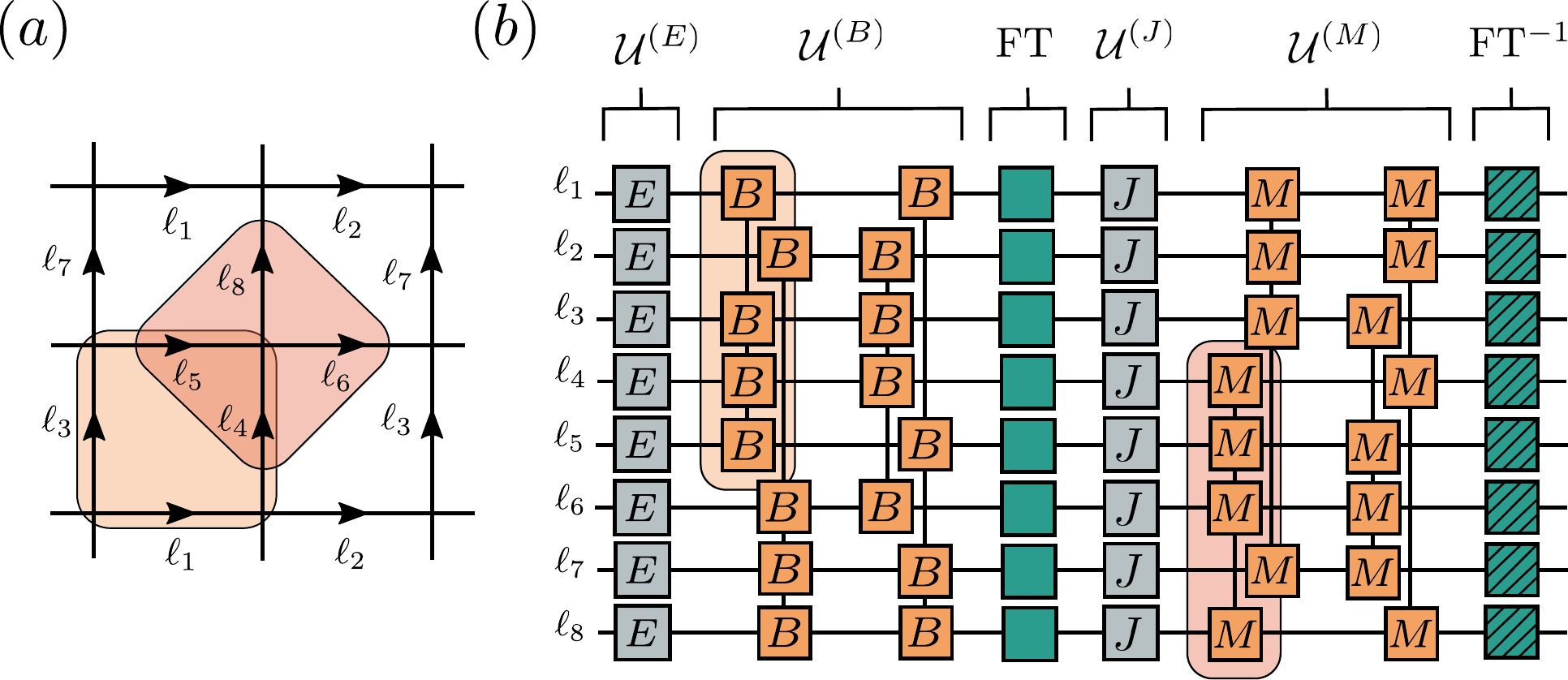}
    \caption{{\bf Abelian-Higgs Trotter step:} (a) Periodic plaquette: a two by two lattice with periodic boundary conditions consists of four different plaquettes and can be simulated using eight qudits. (b) Circuit representation of a single Trotter step to digitally simulate the time evolution of the Abelian-Higgs model on a periodic plaquette. Both the gauge and the matter variables can be addressed with a similar circuit up to a Fourier transform (FT), using single-qudit as well as four-qudit plaquette gates. Two examples of the four-body gates and the links involved are highlighted in (a) and (b) for the magnetic (orange) and matter (red) part of the Hamiltonian.}
    \label{fig:AH_circuit}
\end{figure}

We discuss now the (2+1)D Abelian-Higgs model (AHM)~\cite{Fradkin_1979} -- more precisely U($1$) LGT coupled to a complex Higgs field with frozen radial mode. The AHM is described by the Hamiltonian $H_\text{AHM} = H_{\text{U}(1)} + H'$ with additional ``mass'' ($M$) and ``coupling'' ($J$) terms given by~\cite{Gonzalez_2017}
\begin{align}
     H'=  &\frac{\lambda_M}{2} \sum_+ \Pi_+^2 \nonumber\\ &\quad+ \lambda_J \sum_{\ell=(+,+')} \left(e^{i\phi_+} Q_+ e^{-i\phi_{+'}} + \text{h.c.}\right) \;,
\end{align}
where the matter fields $e^{i\phi_+}$ and $\Pi_+$ live on lattice sites $+$ where neighboring links form a vertex. The coupling between matter and gauge fields is determined by the term $\propto \lambda_J$ that acts on a link $\ell = (+,+')$ connecting two neighboring sites $+$ and $+'$. The matter fields represent another set of conjugate variables that fulfill the U($1$) commutation relation $[\Pi,e^{i\phi}] = e^{i\phi}$, so we can digitize them in the same way as discussed above for the gauge fields, replacing $\Pi^2 /2 \rightarrow -d^2/(2\pi^2) \cos (2\pi\Pi/d)$.

By construction, the combined (digitized) Hamiltonian 
$H_\text{AHM}$ is invariant under $\mathbb{Z}_d$ gauge transformations, i.e. $H_\text{AHM} = \mathcal{V} H_\text{AHM}\mathcal{V}^\dagger$ with $\mathcal{V} = \prod_+ \left[\mathcal{V}_+\right]^{n_+}$ for arbitrary $n_+ = 0, \dots, d-1$. For simplicity, let us focus on a 2D spatial lattice, where the four links $\ell_1, \ell_2, \ell_3, \ell_4$ meet at the vertex $+$, such that the local transformation is given by the Gauss' law operator
\begin{align}
    \mathcal{V}_+ = P_{\ell_1} P_{\ell_2} P^\dagger_{\ell_3} P^\dagger_{\ell_4}e^{-2\pi i \Pi_+/d} \;.
\end{align}

Given a gauge-invariant state, we can solve the corresponding Gauss' law  $\prod_x\mathcal{V}_+|\psi \rangle = |\psi \rangle$ to eliminate the matter degrees of freedom completely from the problem. The result is a $\mathbb{Z}_d$ Toric code-type model of the form $H_{\text{AHM},d} = H_{\mathbb{Z}_d} + H'_d$ with
\begin{align}
     H'_d &=  \lambda_M\sum_{\ell}\left(Q_\ell + Q_\ell
^\dagger\right) \nonumber \\ &\quad-\frac{\lambda_J d^2}{2\pi^2}  \sum_+ \left(P_{\ell_1} P_{\ell_2} P_{\ell_3}^\dagger P_{\ell_4}^\dagger + \text{h.c.}\right)
\end{align}

Let us now come back to the realization of the AHM with qudits. 
The duality between the two bases $\{|\varphi\rangle\}$ and $\{|n\rangle\}$ allows us to switch back and forth using the single-qudit gate that realizes the discrete Fourier transform. Since this essentially exchanges the roles of the operators $P$ and $Q$, the duality also maps $H_{\mathbb{Z}_d}$ onto $H'_d$ and vice versa. We can therefore simulate the real-time dynamics of $H_{\text{AHM}}$ by alternating applications of the Trotter gates parametrized by $\lambda_E$ and $\lambda_B$, and the \emph{same gates} with different parameters $\lambda_E \rightarrow \lambda_J$ and $\lambda_B \rightarrow \lambda_M$, interlaced with alternating forward and backward Fourier transforms (Fig.~\ref{fig:AH_circuit}).

We emphasize that this digitization of the AHM is very resource-efficient. After solving Gauss' law, there is no gauge redundancy left and hence every qudit of information is used in the quantum simulation. We note that, already for $d = 2$, $\mathbb{Z}_d$ LGTs coupled to dynamical matter show interesting properties such as  topological order~\cite{Fradkin_2013, Assaad_2016, Wen_2017, Gonzalez_2020, Borla_2022} and unconventional dynamics~\cite{Iadecola_2020, Sai_2022, Halimeh_2022}.
Moreover, the (2+1)D AHM recovered in the limit $d\rightarrow \infty$ is one of the simplest realistic models that displays dynamical confinement in the presence of matter. We will now use this example to show that our architecture enables to study the real-time dynamics of string-breaking, which presents an outstanding challenge where future quantum simulation might outperform classical supercomputers.

\subsection{Dynamical string breaking}
\label{sec:physics_string_breaking}

\begin{figure}
    \centering
    \includegraphics[width=\columnwidth]{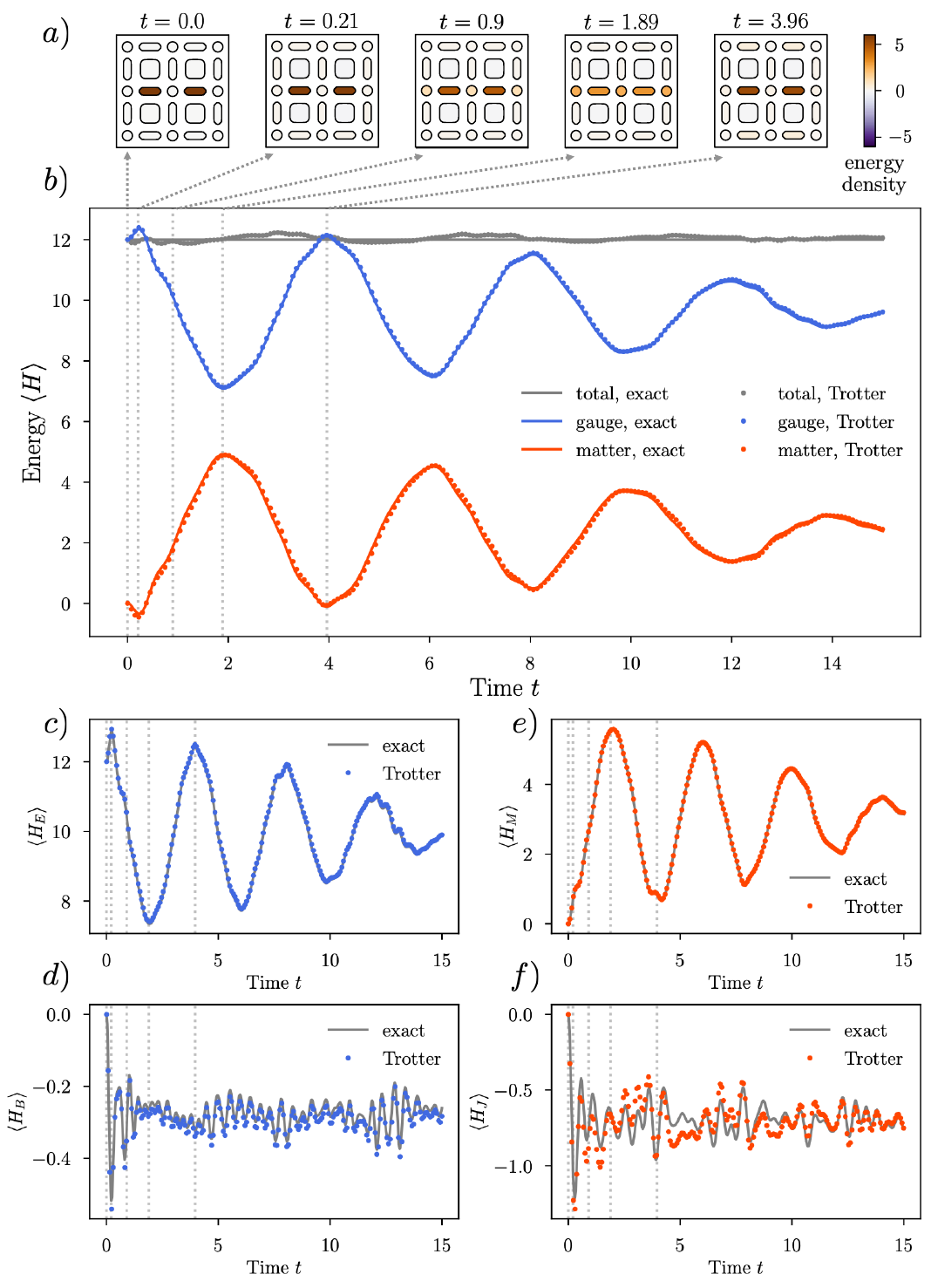}
    \caption{{\bf Particle-pair creation:} We show quench dynamics following an initial flux string on two links of a periodic $2\times 2$ lattice for a $\mathbb{Z}_d\subset$ U($1$) digitization of the AHM with $d=3$. The parameters are 
    $\lambda_E = \frac{4 \pi }{9} \approx 1.4$, $\lambda_B = 0.5$, $\lambda_M = 0.5$ and $\lambda_J = \frac{2 \pi}{9} \approx 0.7 $. $a)$ The pictures show snapshots of the average local electric ($\propto \lambda_E$), magnetic ($\propto \lambda_B$) and matter ($\propto \lambda_M$) energies on the links, plaquettes and sites, respectively. The corresponding times are indicated by light grey dotted lines in panels $b)-f)$, where we show various energies as a function of time. We compared exact results (solid lines) to Trotterized dynamics (small dots) for a second order Trotter decomposition. $b)$ shows the total energy (grey) and its contributions from the gauge (blue) and matter sector (red). The gauge part further contains electric ($\propto \lambda_E$, panel $c)$) and magnetic ($\propto \lambda_B$, panel $d)$) contributions. Similarly, the matter sector is the sum of the contributions shown in $e)$ ($\propto \lambda_M$) and $f)$ ($\propto \lambda_J$).}
    \label{fig:pair_creation}
\end{figure}

One of the most striking features of gauge theories is the confinement of charges. In a pure lattice gauge theory described by the Kogut-Susskind Hamiltonian [Eq.~\eqref{eq:KS_hamiltonian}], confinement is easiest to understand in the strong-coupling limit with $\lambda_B/\lambda_E \rightarrow 0$. A state with two static charges on top of the vacuum is then described by a product state of electric energy eigenstates with $E^2 = 0$ everywhere except on a line of links, where $E^2 = 1$, connecting the locations of the two charges. Consequently, the total energy $\mathcal{E}$ of such a configuration scales like $\mathcal{E} \propto \lambda_E R$, where $R$ is the distance between the two particles. This linear potential gives rise to a constant force, independent of $R$, which confines the charges into a strongly bound pair. For finite $\lambda_B$, and in the continuum limit, the plaquette term complicates this simplistic picture and turns the flux string into an extended object with rich dynamical features. For dynamical instead of static charges, the situation becomes even more interesting: when the energy of the flux tube exceeds the rest mass of the charged matter particles, pairs of particles and anti-particles may be created whose charges can compensate the gauge fields within the flux tube. As a consequence, a single flux tube can tear into two pieces, an effect that is known as string-breaking. Achieving a quantitative understanding of the real-time dynamics of string-breaking constitutes an outstanding challenge, where quantum simulations may help to improve our understanding of fundamental physics.

In this section, we illustrate that our approach will enable to study effects related to string-breaking. For simplicity, we focus on the case of the Abelian-Higgs model discussed above. Due to the limitations of classical benchmark simulations, we restrict ourselves to one of the smallest non-trivial systems, namely a $2 \times 2$ lattice with periodic boundary conditions. For a chosen digitization $\mathbb{Z}_d\subset$ U($1$), this situation corresponds to a register of eight qudits with size $d$. In order to mimic the string-breaking dynamics, we initialize a periodic flux string on two links, which is an eigenstate of the system for $\lambda_B=\lambda_M=\lambda_J=0$ [see $t=0$ in Fig.~\ref{fig:pair_creation}(a)]. This state can be prepared easily by applying appropriate single-qudit rotations on a given trivial product state. We then initiate dynamics by executing the Trotterized time evolution given by the circuit shown in Fig.~\ref{fig:AH_circuit} with fixed parameters $\lambda_B,\lambda_M,\lambda_J \neq 0$. 

In Fig.~\ref{fig:pair_creation}, we present a numerical simulation of the resulting dynamics for a very small qudit size with $d=3$. Fig.~\ref{fig:pair_creation}(a) illustrates the dynamics of the flux string by showing the spatial distribution of the electric, magnetic and matter energy densities over links, plaquettes and sites, respectively. Even though the string does not break due to the small system size, its energy clearly decreases while the energy on lattice sites simultaneously increases. This process is the analog of particle-anti-particle pair creation in the full model. We further observe a partial reversal of this dynamics, leading to pronounced oscillations of the energy transfer between gauge and matter sectors [Fig. ~\ref{fig:pair_creation}(b)], reminiscent of the plasma oscillations known to follow Schwinger pair production~\cite{Hebenstreit_2013, Martinez_2016}.

A closer inspection of the four contributions to the total energy [Fig.~\ref{fig:pair_creation}(c)-(f)], reveals two distinct time scales present in this particular quench. To resolve also the smaller time scale reasonably well [see Fig.~\ref{fig:pair_creation}(d) and (f)], we have chosen a relatively small Trotter step. This choice resolves the slower oscillations seen in Fig.~\ref{fig:pair_creation}(a) with the shown accuracy, conserving the total energy with a maximum error of $\delta \langle H \rangle \approx 0.22$. This accuracy requires about $\gtrsim 55$ Trotter steps to reach the time $t\gtrsim 4$ of a single ``plasma'' oscillation. For a simple second order Trotter decomposition, a circuit with a depth of $6$ general single-qudit gates, $4$ diagonal single-qudit gates and $12$ group-multiplication two-qudit gates is sufficient to implement a single Trotter step, independent of system size. In Ref.~\cite{Gonzalez_2022}, we showed how similar requirements can be satisfied in the Rydberg architecture using realistic experimental estimates qudits of size $d = 8$.

\begin{figure}
    \centering
    \includegraphics[width=\columnwidth]{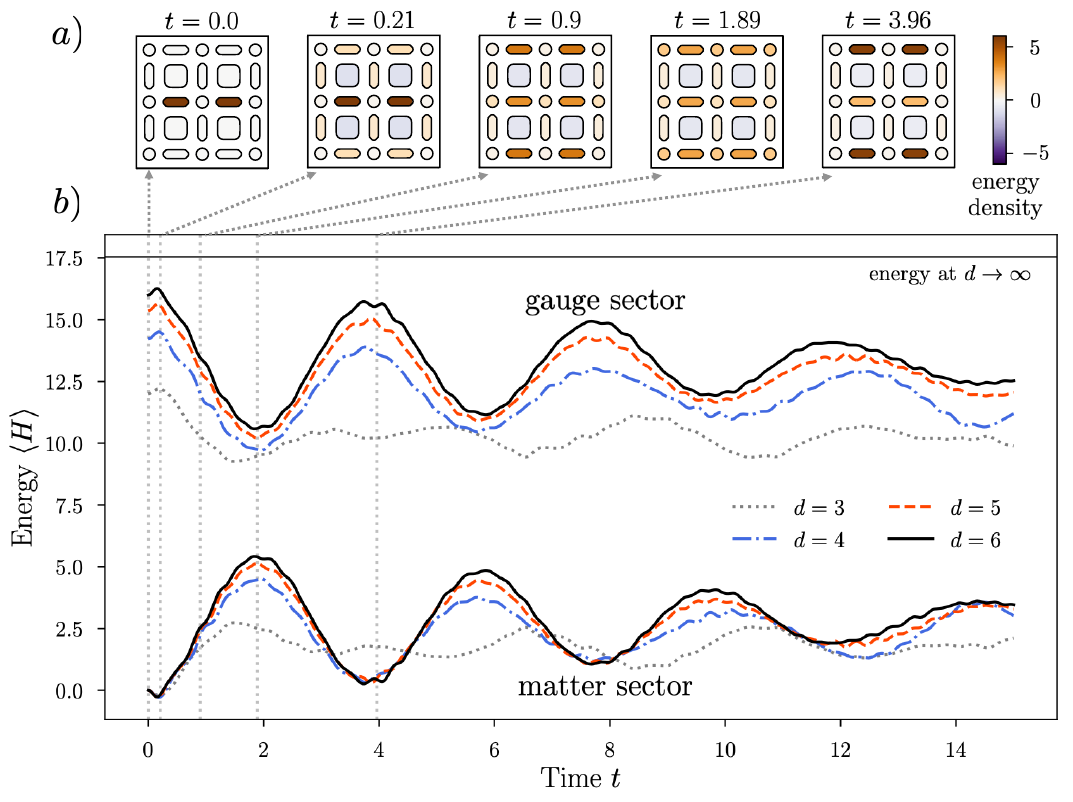}
    \caption{{\bf Scaling to U($1$):} We show the analogous quench dynamics with the same parameters as in Fig.~\ref{fig:pair_creation} for $d=3$, except now $\lambda_B = 2$. $a)$ illustrates the evolution of the local energies, here for $d=6$, analogous to Fig.~\ref{fig:pair_creation}a. $b)$ shows the total energy in gauge and matter sectors, as in Fig.~\ref{fig:pair_creation}b, for several values of $d=3,4,5,6$. Note that the total energy of the initial state depends on $d$ and converges to the corresponding $d\rightarrow \infty$ value of the U($1$) theory, which is indicated by the horizontal black line at the top of the plot.}
    \label{fig:d_extrapolation}
\end{figure}

The above example illustrates the possibility to observe pair production dynamics, an ubiquitous phenomenon in gauge theories, with our proposal. Ultimately, we are of course not interested in the crude lattice regularization with $d=3$, but in the continuum  limit $d\rightarrow\infty$. To this end, we present another set of exact (not Trotterized) simulation results in Fig.~\ref{fig:d_extrapolation}. Here, we have taken the same system size and quench setup as discussed above, but increased the strength of the plaquette interaction $\lambda_B$. Our choice mimics the approach to the continuum limit, where the plaquette term becomes more important, which also broadens the electric flux string. The impact on the dynamics is clearly visible in Fig.~\ref{fig:d_extrapolation}(a): the pair production dynamics is now super-imposed with transverse fluctuations of the flux string, which also builds up significant magnetic field fluctuations on the neighboring plaquettes.
In these simulations, we have also systematically increased the size $d$ of the qudit digitization of the fields. As illustrated with the energy transfer between gauge and matter sectors shown in Fig.~\ref{fig:d_extrapolation}(b), we observe a relatively fast convergence for $d\leq 6$. These results suggest that the continuum U$(1)$ dynamics of the Abelian-Higgs model can be well approximated using our qudit protocol, where each local gauge field is enconded into a single Rydberg atom.

\section{Non-abelian hadronic matter}
\label{sec:non_abelian_hadrons}

Having discussed the abelian case in detail, we now turn to non-abelian gauge groups. Our main interest are Yang-Mills theories with gauge groups SU($N$), the simplest example being SU($2$). 
In Sec.~\ref{sec:SU(2)_lgt}, we specify the requirements of the qudit protocol to simulate finite subgroups of SU($2$). In Sec.~\ref{sec:dynamical_structure}, we show how to prepare hadronic states on a fermion-qudit processor using both adiabatic as well as variational techniques, and how to investigate their internal structure by measuring real-time correlation functions.

\subsection{SU($2$) lattice gauge theory}
\label{sec:SU(2)_lgt}

For SU($2$), the link operators $U$ are $2\times2$ matrices defined by their action
\begin{align}
    U_{ab}|g \rangle = D_{ab}(g) |g \rangle
\end{align}
on group element states $g$. Here $D(g)$ denotes the fundamental matrix representation of $g\in$ SU($2$), which is the corresponding unitary $2\times2$ matrix with unit determinant. Consequently, the action of the plaquette operator becomes
\begin{align}
    &\mathcal{U}_\square |g_1\rangle |g_2\rangle |g_3\rangle |g_4\rangle  \nonumber\\ &\quad= \text{tr} \left[D(g_1g_2 g_3^{-1}g_4^{-1})\right] |g_1\rangle |g_2\rangle |g_3\rangle |g_4\rangle  \;,
\end{align}
where the trace $\text{tr}[D] = \sum_a D_{aa} = \chi$ defines the character $\chi$ of the representation.

Similar to the abelian case, we also have a dual basis spanned by states $|jmn\rangle$ where $j = 0,1/2, 1, \dots$ labels all irreducible representations (irreps) of SU($2$) and $m,n = -j, -j+1, \dots j$. Explicitly they are given by
\begin{align}
    |jmn\rangle = \sqrt{(2j+1)}\int dg D^{(j)}_{mn}(g) | g\rangle \;,
\end{align}
where $dg$ denotes the Haar measure on SU($2$) and $D^{(j)}$ denote the different irreps.
The electric energy operator $E^2$ is defined as the quadratic Casimir of SU($2$), i.e. it acts as
\begin{align}
    E^2 |jmn\rangle = j(j+1) |jmn\rangle \;.
\end{align}
With these definitions of $E$ and $U$ for every link, the full dynamics of pure SU($2$) LGT is given again by the Hamiltonian in Eq.~\eqref{eq:KS_hamiltonian}.

As a matter of fact, there are only a finite number of finite non-abelian subgroups of SU($2$). These are the quaternion group $Q_8$ and the binary tetrahedral, octahedral and icosahedral groups $2T$, $2O$ and $2I$ with group sizes $8,24,48$ and $120$, respectively. Nevertheless, it has been shown that the largest two groups can provide a useful approximation of the desired low-energy physics~\cite{Petcher_1980, Alexandru_2019}. The simplest digitization of SU($2$) is given by its smallest non-abelian subgroup $Q_8$, the quaternion group, an example that is worked out in detail in Ref.~\cite{Gonzalez_2022}. $Q_8$ consists of the eight elements $\{\mathbf{1}, -\mathbf{1}, i\sigma^z, -i\sigma^z, i \sigma^y, -i \sigma^y, i \sigma^z, -i \sigma^z\}$ with Pauli matrices $\sigma^j$ and $\mathbf{1}$ the identity matrix, which we assign to eight qudit states $\{|0\rangle, \dots |7\rangle\}$. In this basis, group multiplication is represented by $8\times 8$ permutation matrices.

The digitization of the magnetic interaction proceeds completely analogous to the abelian case. Restricting to the chosen subgroup, we obtain
\begin{align}
    f_B(g) = e^{-2i\lambda_B \chi(g)\delta t } \;,
\end{align}
with the character $\chi(g)$ in the fundamental representation. For example, for $Q_8$ the only non-trivial contribution comes from $\chi(\pm \mathbf{1}) = \pm 2$, otherwise $\chi=0$. This determines the single-qudit phase gate required for the plaquette term. Again, we additionally need controlled two-qudit gates $\Theta(\ell|\ell')$ that realize group multiplication, assembled for $Q_8$ from the eight permutation matrices specified in Ref.~\cite{Gonzalez_2022}.

The digitization of the electric term is not unique. The problem lies in the fact that there exists no analog of the quadratic Casimir for finite groups. One generally available option is to define the Hamiltonian instead through a lattice path integral~\cite{Lamm_2019}. This leads to an expression of the form
\begin{align}
    f_E(g',g) = \left[\exp \left(i \delta t \log T_E \right)\right]_{g',g} \;, 
\end{align}
involving a matrix logarithm of the Euclidean (imaginary-time) electric transfer matrix $T_E$ and a subsequent re-exponentiation to obtain a unitary operator for the real-time dynamics. For SU($2$), $T_E$ is given by
\begin{align}
    \left[T_E\right]_{g',g} = e^{(2/\lambda_E) \chi(g'g^{-1})} \;.
\end{align}
Further details are provided in Ref.~\cite{Gonzalez_2022} for the $Q_8$ digitization. 
In any case, the function $f_E$ can be calculated efficiently classically, which determines the required one-qudit gate that realizes the electric part of the dynamics.

This example of SU($2$) demonstrates that the complexity of simulating the dynamics of non-abelian LGTs is identical to the abelian case in the sense that both require the implementation of general single-qudit gates and controlled two-qudit gates that realize group multiplication. Consequently, a modest qudit size of $d=8$ can be employed for approximating either an abelian U($1$) LGT via the subgroup $\mathbb{Z}_8$ or a non-abelian SU($2$) LGT via the subgroup $Q_8$.

\subsection{Dynamical structure of hadrons}
\label{sec:dynamical_structure}

As a simple toy model to illustrate our ideas, we consider the $Q_8$ truncation of SU(2) lattice gauge theory with staggered fermions in the fundamental representation. To simplify the numerical benchmark, we focus on a one-dimensional chain with (anti-)periodic boundary conditions for the (fermions)bosons. As the local electric field energy, we choose an operator $E^2 = \sum_{jmn} E^2_j |jmn\rangle \langle jmn |$ in the representation basis, with $E_0 = 0, E_{1/2} = 3/4$ and $E_{I/J/K} = \infty$, corresponding to the smallest non-trivial truncation of the SU(2) gauge fields. To summarize, in lattice units ($\bar{H}=2/(ag^2) H$ with lattice spacing $a$ and gauge coupling $g^2$), we consider the rescaled Hamiltonian~\cite{hamer1977lattice}
\begin{align}
    \bar{H} &= \sum_n \left(E_{n,n+1}^2 + \mu (-1)^n \psi_n^\dagger \psi_n \right) \nonumber \\
    &\overset{(+)}{-}ix \sum_n \left(\psi_n^\dagger U_{n,n+1} \psi_{n+1} - \text{h.c.} \right) \;.
\end{align}
Here, $\mu = 2m/(g^2 a)$ is related to the bare fermion mass $m$ and $x = 1/(g^2 a^2)$ controls the gauge-matter coupling ($x\rightarrow\infty$ in the continuum limit). $\psi_n$ are two-component fermionic operators located at every lattice site $n$, $U_{n,n+1}$ are the ($Q_8$) gauge field operators on the links between two-lattice sites and the $+$ sign is taken for exactly one hopping term on an $N$-site chain.

\subsubsection{Preparing hadrons}
A first useful application of our approach is the preparation of bound states of matter (hadrons) that consist of multiple elementary particles. Here, the generic approach is (a)diabatic state preparation with real-time Trotter steps starting from a trivial product state with the right quantum numbers (such as total charge, baryon number, etc). An alternative is variational state preparation, where the availability of unitary time evolution steps corresponding the underlying Hamiltonian, as provided by our approach, can be helpful as well.

As an example, we consider the lowest-lying baryon of $\bar{H}$. To identify this state, note that $\bar{H}$ commutes with the operator $N_b = \frac{1}{2}\sum_n \left(\psi_n^\dagger \psi_n - N\right)$, which counts the total baryon number. In the limit $x\rightarrow 0$, we can thus identify a baryon with total momentum $p$ as the product state without gauge flux ($|jmn = 000\rangle$) and an appropriate fermion filling, i.e.
\begin{align}
    |B;p \rangle &\propto \left[\bigotimes_{(n,n+1)} | 0 0 0 \rangle \right] \nonumber\\ &\quad\otimes \left[\bigotimes_n e^{2\pi i pn/N} \psi_{n,1}^\dagger \psi_{n,2}^\dagger |O\rangle \right] \;,
\end{align}
where $|O\rangle$ is the fermionic ground state of the staggered fermion mass term at half-filling with $N_b=0$ and energy $2N\mu$. The baryon $|B;p\rangle$ is the ground state in the sector with $N_b=1$ and total momentum $p$ with energy $2\mu$ above the vacuum. The product state written above can be prepared more-or-less directly in an experiment, and the more complicated entangled state at finite $x$ can be subsequently reached by an appropriate application of finite Trotter steps.

\begin{figure}[t]
    \centering
    \includegraphics[width=\columnwidth]{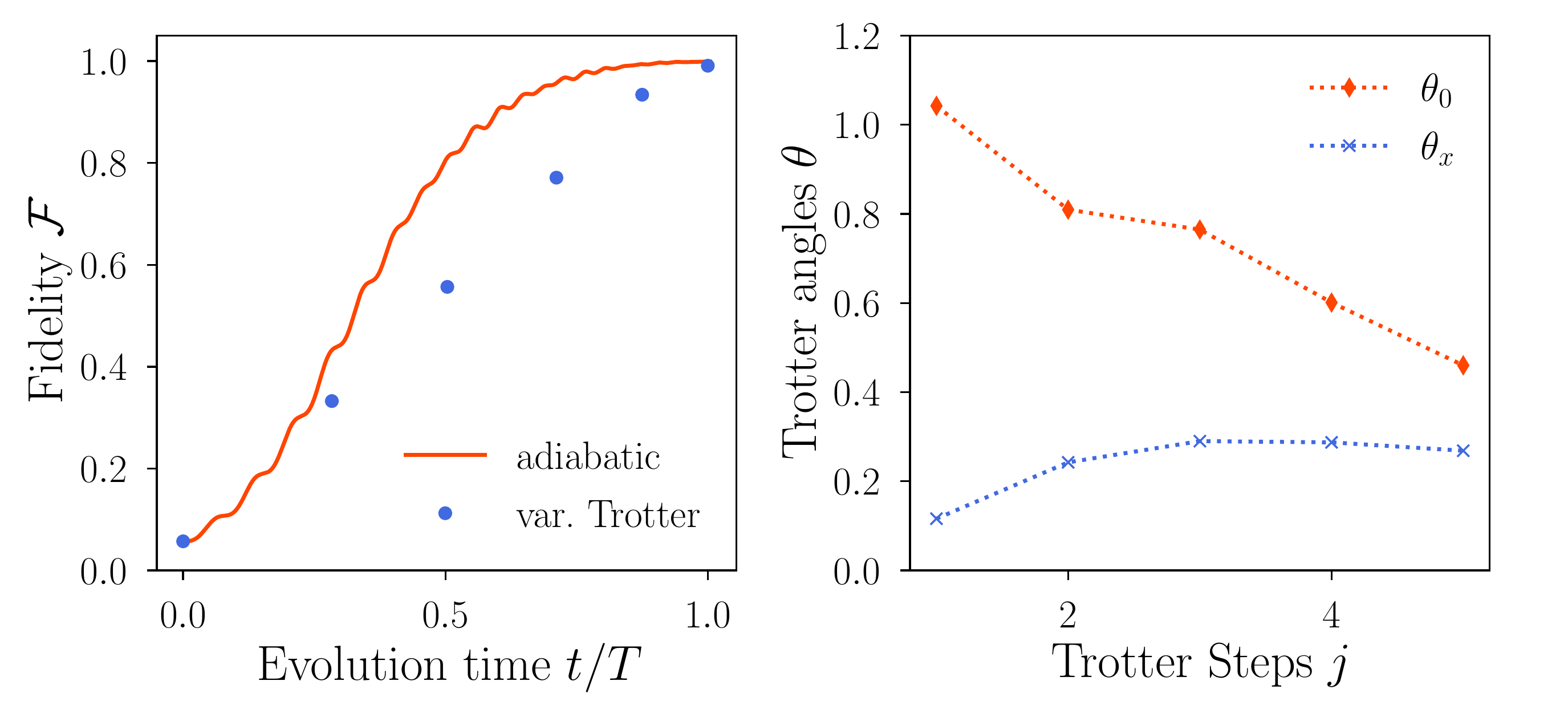}
    \caption{{\bf Baryon preparation:} Left: Fidelity of the adiabatic and a variational state preparation protocol. Right: optimized variational parameters for the Trotterized state preparation, which achieves a fidelity of $\mathcal{F}\approx 99.09\%$ with five steps of a second-order Trotter decomposition of the form $e^{-i \theta_0 H_0/2} e^{-i \theta_x H_x} e^{-i \theta_0 H_0/2}$. }
    \label{fig:state preparation}
\end{figure}

To illustrate these ideas, we show in Fig.~\ref{fig:state preparation} the results of a numerical simulation for a chain of $N=8$ sites. As expected the target baryon can be prepared adiabatically. Remarkably, we find that also an optimized circuit consisting of a few Trotter steps with a few variational parameters is sufficient for state preparation with high fidelity.
For the present example, a single Trotter step corresponds to a circuit of depth four consisting of two parallelized layers of the gauge-matter interaction, each involving two group multiplication gates (see section \ref{sec:fermionic_matter_fields}). Assuming a finite gate fidelity of $\approx 99.6\%$ per group multiplication gate~\cite{Gonzalez_2022}, we thus estimate an achievable fidelity for a single Trotter step of $\gtrsim 99\%$, comparable to the Trotter error in Fig.~\ref{fig:state preparation}. From this estimate, we conclude that our approach is accessible for preparing the desired baryon with near-term quantum hardware.

\subsubsection{Probing the hadronic structure}
Understanding the internal structure of hadrons constitutes an outstanding challenge for lattice gauge theory calculations due to the complication of evaluating real-time correlation functions. One important object in this context, is the so-called hadronic tensor, which encodes the information of various structure functions and is directly relevant for deeply-inelastic scattering that is traditionally used to probe the partonic structure of hadrons. As proposed in~\cite{lamm2020parton}, we thus focus on the hadronic tensor in the following (see, e.g., \cite{liang2020towards} for  an extraction of the hadronic tensor from lattice QCD).

The hadronic tensor is defined as the time-ordered two-point correlation function of gauge-invariant fermion currents,
\begin{align}
    W^{\mu \nu}(q|p) = \text{Re} \int dx\, e^{-iqx} \langle p | j^\mu(x) j^\nu(0) | p \rangle \;. 
\end{align}
Here, $|p\rangle$ is a hadronic state with momentum $p$, $q$ is a space-time momentum corresponding to the space-time coordinate $x$ and $j^\mu = \psi^\dagger \gamma^0 \gamma^\mu \psi$ is the fermion current operator. In our case for the $Q_8$ baryon with staggered fermions, we can define an analogous quantity on the 1D lattice as
\begin{align}
  &W^{\mu \nu}(k,w|B;p) \nonumber\\
  &\quad= \sum_{x=0}^{N/2} \int_{-\infty}^\infty dt \, e^{-i (wt + 2\pi kn/N)} W^{\mu \nu}(x,t|B;p),
\end{align}    
where 
\begin{align}
W^{\mu \nu}(x,t|B;p) = \text{Re} \left[\langle B;p  | e^{iHt} j^\mu_x e^{-iHt} j^\nu_0|B;p  \rangle \right]
\end{align}
is the real-time lattice correlator. Note that the discrete sum runs over $N/2$ lattice sites because of the staggered fermion discretization. The current operators with $\mu,\nu \in \{0,1\}$ at super-site $x$ corresponding to the lattice sites $2n,2n+1$ are
\begin{align}
    j^{0}_x &= \psi^\dagger_{2n}\psi_{2n} + \psi^\dagger_{2n+1}\psi_{2n+1} \;,\\
    j^{1}_x &= \psi^\dagger_{2n} U_{2n,2n+1}\psi_{2n+1} + \text{h.c.} \;.
\end{align}

\begin{figure}[t]
    \centering
    \includegraphics[width=\columnwidth]{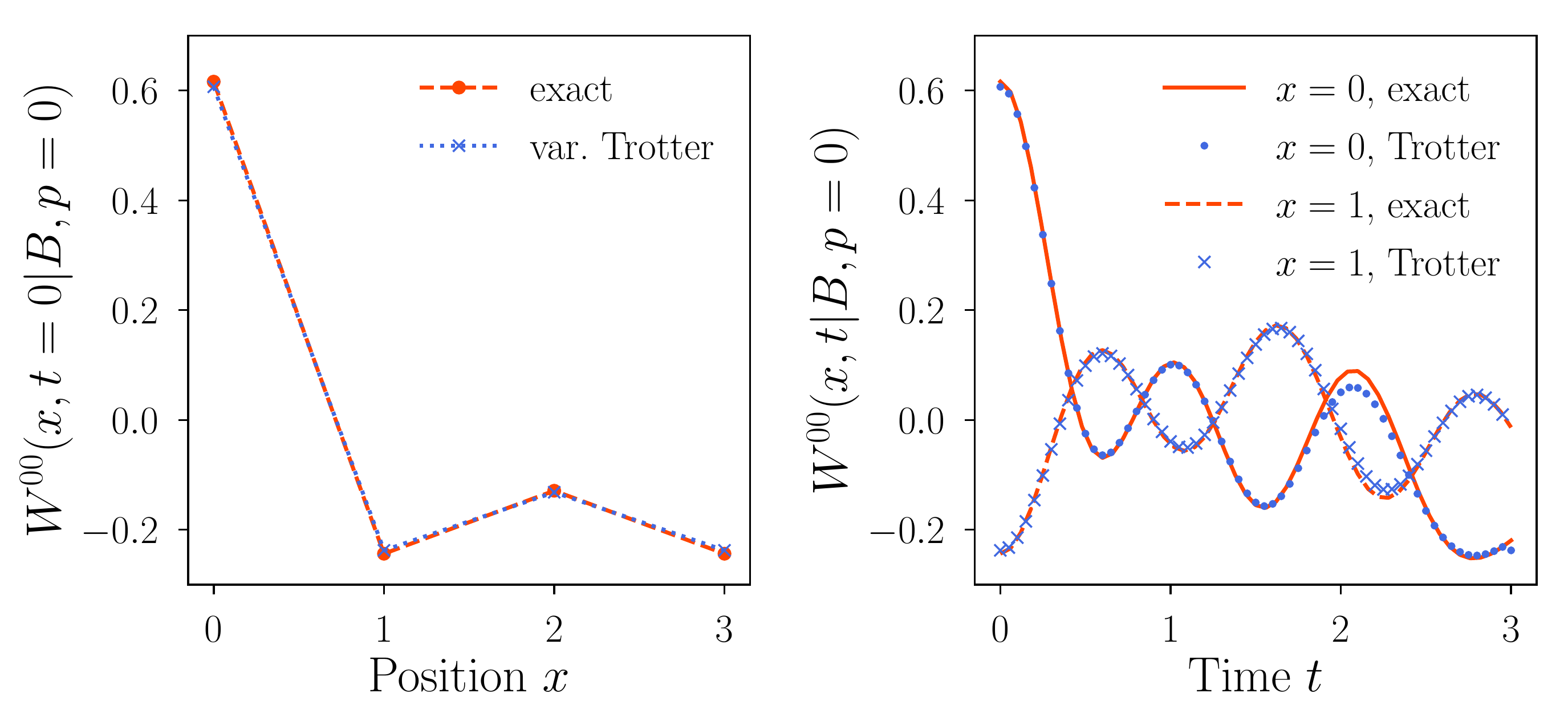}
    \caption{{\bf Baryonic structure:} Density-density component ($\mu=\nu=0$) of the Hadronic tensor $W^{\mu\nu}(x,t|B,p)$ for the baryon $B$ at zero momentum ($p=0$). Left: For $t=0$, the variational state preparation (see figure~\ref{fig:state preparation}) of the baryon provides excellent agreement with the exact result. Right: For $t>0$, the hadronic tensor obtained by an additional Trotterized time evolution also agrees with well with the exact result for the shown Trotter step size.}
    \label{fig:hadronic tensor}
\end{figure}

For simplicity, we focus on the density-density component ($W^{00}$) for the baryon at zero momentum $p=0$ in the following. As demonstrated in Fig.~\ref{fig:hadronic tensor}, the approximate state preparation accurately captures the functional form of $W$ at the initial time $t=0$. Evaluating $W$ at later times requires (neglecting the measurement protocol for the moment) real-time evolution, which we again realize with second-order Trotter steps. As shown in Fig.~\ref{fig:hadronic tensor}, we find good agreement between the exact result and the Trotterized evolution, where a few tens of Trotter steps are sufficient to resolve the first few oscillations.
We emphasize that the only necessary entangling operations here are the group-multplication gates required to decompose the gauge-matter interactions, with a total number of $2N$ per Trotter step. For a gate fidelity of $\approx 99.6\%$~\cite{Gonzalez_2022} for these entangling gates, we estimate that $\mathcal{O}(10)$ Trotter steps are possible on current hardware before the overall fidelity decreases below $\lesssim 90\%$.

\section{Conclusion and outlook}

In this work, we presented a fermion-qudit quantum processor based on Rydberg atoms trapped in optical tweezers, tailored to quantum simulate lattice gauge theories in a hardware-efficient manner. Our proposal is based on a local encoding of both gauge fields and dynamical matter, either fermionic or bosonic, using multi-level atoms that satisfy the proper particle statistics. The latter leads to a circuit decomposition to simulate Trotter dynamics in terms of native atomic gates that retain the locality of the simulated interactions, therefore reducing the required experimental resources to towards a practical quantum advantage. We estimated these resources for two paradigmatic examples relevant to particle physics, including the simulation of dynamical string breaking and the calculation of the structure of non-abelian hadrons, showing how our protocol could be implemented using near-term quantum devices.

Our current formulation is designed for finite groups. One of the challenges of our approach lies in approximating continuous non-abelian groups since they only possess a finite number of finite subgroups. Although in many cases the largest subgroup is sufficient to properly describe the physics in the continuum limit, at least in equilibrium~\cite{Alexandru_2019}, this is still an open question in the non-equilibrium case, where the corresponding approximation might not be good enough in certain regimes. We stress, however, that the quantum hardware presented in this work is also suitable to run different quantum algorithms to simulate gauge-theory dynamics than the one presented here. In particular, we could choose different ways of truncating the infinite Hilbert space of local gauge fields to avoid the subgroup limitation described above, in order to obtain a controlled approximation to continuous non-abelian groups. We will address this in future work~\cite{zache2023quantum}.

\paragraph*{Acknowledgements.--} This work was supported by
the US Air Force Office of Scientific Research (AFOSR) via IOE Grant No. FA9550-19-1-7044 LASCEM, the European Union’s Horizon 2020 research and innovation program under Grant Agreement No. 817482 (PASQuanS), and by the Simons Collaboration on Ultra-Quantum Matter, which is a grant from the Simons Foundation (651440, P.Z.).

\bibliographystyle{quantum}
\bibliography{bibliography}

\end{document}